\documentclass[twocolumn,amsmath,amssymb,aps,superscriptaddress,longbibliography,10pt]{revtex4-2}

\usepackage{graphicx}
\usepackage{dcolumn}
\usepackage{bm}
\usepackage{color}

\usepackage{url}
\usepackage{braket}
\usepackage{comment}
\usepackage{here}
\usepackage{algorithm}
\usepackage{algpseudocode}
\usepackage{booktabs}
\usepackage[colorlinks=true,citecolor=blue,linkcolor=blue,urlcolor=blue]{hyperref}

\begin{document}
\title{Neural-network-assisted Monte Carlo sampling trained by Quantum Approximate Optimization Algorithm}

\author{Yuichiro Nakano}
\email{u830977g@ecs.osaka-u.ac.jp}
\affiliation{
    Graduate School of Engineering Science, The University of Osaka, 1-3 Machikaneyama, Toyonaka, Osaka 560-8531, Japan.
}
\author{Ken N. Okada}
\email{okada.ken.qiqb@osaka-u.ac.jp}
\affiliation{
    Center for Quantum Information and Quantum Biology, The University of Osaka, 1-2 Machikaneyama, Toyonaka, 560-0043, Japan.
}
\author{Keisuke Fujii}
\email{fujii.keisuke.es@osaka-u.ac.jp}
\affiliation{
    Graduate School of Engineering Science, The University of Osaka, 1-3 Machikaneyama, Toyonaka, Osaka 560-8531, Japan.
}
\affiliation{
    Center for Quantum Information and Quantum Biology, The University of Osaka, 1-2 Machikaneyama, Toyonaka, 560-0043, Japan.
}
\affiliation{
    Center for Quantum Computing, RIKEN, Wako Saitama 351-0198, Japan.
}

\date{\today}

\begin{abstract}
Sampling problems are widely regarded as the task for which quantum computers can most readily provide a quantum advantage. 
Leveraging this feature, the quantum-enhanced Markov chain Monte Carlo [Layden, D. \textit{et al.}, \href{https://doi.org/10.1038/s41586-023-06095-4}{Nature 619, 282–287 (2023)}] has been proposed recently, where sampling from a quantum computer is used as a proposal distribution and convergence to a target distribution is accelerated.
However, guaranteeing convergence to the target distribution typically forces one to impose restrictive symmetry constraints on the quantum circuit, which makes it hard to design good proposal distributions and prevents making full use of the advantage of a quantum computer.
We explore a hybrid quantum-classical MCMC framework that combines a quantum circuit with a generative neural sampler (GNS).
The GNS is trained on quantum samples and acts as a classical surrogate to efficiently emulate quantum outputs, thereby lifting circuit constraints.
We apply this method to Boltzmann sampling of spin glasses using proposals trained with a QAOA circuit. This approach outperforms conventional methods, showing a $\sim$100× improvement in spectral gap over uniform proposals.
Notably, it maintains similar acceleration even without parameter optimization.
These results establish the method as a viable sampling-based quantum algorithm for NISQ devices and highlight its potential for solving practical problems with quantum computation.

\end{abstract}

\maketitle


\section{Introduction}
Quantum computers are widely anticipated as the next generation of computing devices, as they are capable of solving certain problems faster than known algorithms on classical computers \cite{shor1994algorithms, grover1997quantum, abrams1999quantum, PhysRevLett.103.150502}. 
However, current Noisy Intermediate-Scale Quantum (NISQ) devices \cite{preskill2018quantum} lack the capability to implement full quantum error correction, which makes it difficult to execute these algorithms with theoretically proven quantum speedup. 
Still, these devices possess the potential to outperform classical computation, as experimentally demonstrated by the achievement of ``quantum computational supremacy'' \cite{arute2019quantum, doi:10.1126/science.abe8770, Madsen2022}, and active efforts have been made to leverage this capability for practical tasks.
Variational Quantum Algorithms (VQAs) \cite{cerezo2021variational} are the leading strategies to obtain a quantum advantage on NISQ devices.
These methods heuristically find solutions by optimizing objective functions based on results from parameterized shallow quantum circuits and then updating parameters using classical optimization methods.
For example, Quantum Approximate Optimization Algorithm (QAOA) \cite{farhi2014quantum} is one of the most promising VQAs for solving combinatorial optimization problems by searching the ground state of an Ising model that encodes the problem.
While VQAs have achieved significant success in the NISQ era \cite{peruzzo2014variational, farhi2014quantum, farhi2018classification, mitarai2018quantum}, most methods that employ expectation values of Hamiltonian as the objective function remain impractical for real-world applications. 
This is primarily due to the large number of measurements required as circuit size and noise increase \cite{gonthier2022measurements}, as well as the existence of barren plateaus in the optimization \cite{mcclean2018barren}.

The fundamental source of quantum advantage lies in its ability to generate states that are difficult to produce classically, by operating in a Hilbert space that grows exponentially with system size, and to sample classical bitstrings by measuring them.
Indeed, random circuit sampling \cite{hangleiter2023computational}, which is used to demonstrate the quantum supremacy \cite{aaronson2016complexity, arute2019quantum}, relies on these facts. 
On the other hand, if one is only interested in evaluating expectation values, 
then even for large-scale quantum circuits, classical simulation can be performed efficiently as long as the spread of correlations remains limited~\cite{Kim2023, KECHEDZHI2024431}.
This observation suggests a promising usage of NISQ devices as a sampler to generate certain probability distributions, instead of an estimator of expectation values.
With this in mind, a hybrid quantum-classical algorithm known as Quantum-Selected Configuration Interaction (QSCI) has recently been proposed~\cite{kanno2023quantumselectedconfigurationinteractionclassical}
and large-scale demonstrations using quantum hardware have already been conducted~\cite{robledomoreno2024chemistryexactsolutionsquantumcentric}.

Another approach in this direction is the hybrid quantum-classical MCMC \cite{layden2023quantum}, which uses a quantum circuit as the proposal distribution in Markov chain Monte Carlo (MCMC) \cite{metropolis1953equation,hastings1970monte}.
Sampling from the quantum circuit produces new transitions for the Markov chain, and the accept-reject step is performed classically based on the acceptance probability between transitions.
The key idea is to take advantage of quantum circuit sampling without explicitly calculating its output distribution, which typically requires exponential time as system size increases.
Existing works implement the idea by selecting a symmetric circuit $U = U^{\top}$, which can avoid the calculation of output distribution entirely.
However, this complicates the understanding of how the circuit influences the acceptance probability, making it difficult to control the convergence rate \cite{PhysRevResearch.6.033105, christmann2024quantum, PhysRevA.110.052414}.

In this study, we investigate a hybrid quantum-classical MCMC that integrates a quantum circuit sampling with a generative neural sampler (GNS) \cite{pmlr-v15-larochelle11a, uria2016neural, germain2015made} to overcome existing limitations in these approaches. 
Unlike existing methods that impose symmetry constraints on the quantum circuit to avoid calculating output distributions, this approach uses a GNS as a neural surrogate model that approximates the output distribution in a computationally tractable form.
This eliminates the need for restrictive circuit structures and enables the use of arbitrary quantum circuits as proposal distributions. 
Consequently, this work expands the applicability of the hybrid quantum-classical MCMC to a broader class of target distributions while retaining compatibility with NISQ devices.
By enabling flexible and efficient quantum sampling, this framework paves the way for practical applications in domains such as statistical physics, machine learning, and combinatorial optimization.

We focus on Boltzmann sampling for the Ising model using a proposal distribution generated by a QAOA circuit, which does not satisfy the symmetry constraint and thus lies outside the scope of existing methods.
It has been shown that the output distributions of shallow QAOA circuits can approximate low-temperature Boltzmann distributions that include contributions from several excited states \cite{lotshaw2023approximate, diez2023quantum, PhysRevResearch.6.013071}.
A GNS trained on such QAOA outputs can induce transitions weighted toward low-temperature Boltzmann distributions, where long convergence times are typically a bottleneck, thereby enabling more efficient MCMC simulations.
Moreover, shallow QAOA circuits are well suited for implementation on NISQ devices, while sampling from them remains intractable for classical computers \cite{farhi2019quantumsupremacyquantumapproximate}.
We demonstrate the performance of the proposed method on the Boltzmann distribution of fully-connected spin glasses.
The spectral gap of the proposed method remains superior to that of the classical proposals as the system size increases. 
The proposed method improves the spectral gap by approximately two orders of magnitude compared to the uniform distribution, particularly at low temperatures ($\beta \geq 5$), when $n \geq 10$. 
In magnetization estimation, we show the method constructs a proposal distribution that approximates the Boltzmann distribution at low temperatures, allowing transitions in complex energy landscapes where SSF struggles and ensuring unbiased sampling.
In particular, the use of well-chosen initial parameters, known as “fixed angles” \cite{brandao2018fixedcontrolparametersquantum, 9605328, wurtz2021fixedangleconjectureqaoa, 10.1145/3584706}, can obviate the need for parameter optimization, allowing the method to be deployed without costly variational procedures.
Through the numerical experiments, the proposed method by fixed-angle QAOA also surpasses the classical proposals.
This feature may facilitate practical applications that leverage the computational power of NISQ devices with a realistic number of circuit executions.

The remainder of this paper is organized as follows. 
In Sec.~\ref{sec:preliminary}, we introduce the Boltzmann sampling task for spin glasses and provide an overview of the three main components of this study: MCMC, GNS, and QAOA.
Section~\ref{sec:nmcmc_qaoa} provides a hybrid quantum-classical MCMC algorithm based on these three components. 
Section~\ref{sec:numerical_experiment} presents numerical experiments demonstrating that the proposed sampler improves convergence compared to existing methods and offers several additional advantages. 
Finally, Sec.~\ref{sec:conclusion} concludes the paper and discusses future directions.

\section{Background}
\label{sec:preliminary}
\begin{figure*}
    \centering
    \includegraphics[width=0.9\linewidth]{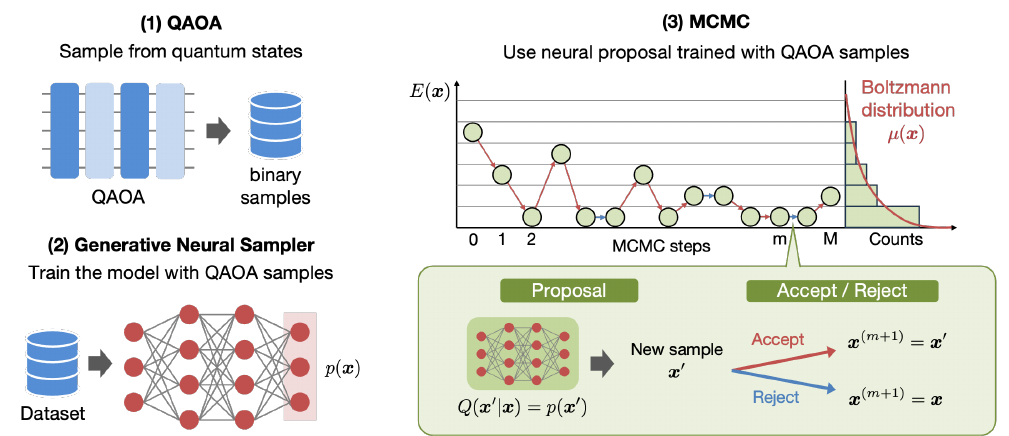}
    \caption{A schematic diagram of the proposed method in this study. The method comprises three main components—QAOA, GNS, and MCMC—and proceeds as follows:
    \textbf{(i)} Run QAOA and sample the output states of the quantum circuit.
    \textbf{(ii)} Use the sampled binary sequences as training data to train a generative neural network that reproduces the QAOA output distribution.
    \textbf{(iii)} Perform MCMC using the trained neural network as the proposal distribution.}
    \label{fig:main_scheme}
\end{figure*}

\subsection{Spin glass and Boltzmann distribution}
\label{sec:spin_glass}
Disordered Ising models with frustration, commonly known as random spin models, have greatly contributed to our understanding of various physical phenomena \cite{doi:10.1142/1655}. 
However, calculating physical quantities for these models poses a significant challenge in classical computation. One example of such a random spin model is the spin glass, which serves as a theoretical framework for describing the glassy phase of spins \cite{RevModPhys.58.801}. A spin glass with $n$ spin variables $x = \{1, -1\}$ is defined by the following energy function:
\begin{align}
    \label{eq:spin_glass}
    E(\bm{x}) = -\sum_{(j,k)} J_{jk} x_j x_k,
\end{align}
where $\bm{x} = {(x_1, \cdots, x_n)}^\top$ denotes the configuration of all spins, and 
$\{J_{jk}\}$ are the interaction strengths between connected spins $(j,k)$, randomly drawn from a standard normal distribution.

To investigate the thermodynamic properties of such physical models, one typically computes the expectation values of physical observables with respect to the Boltzmann distribution. The Boltzmann distribution is a fundamental probability distribution that characterizes the energy states of a physical system at temperature $T$ and is defined as:
\begin{align}
\mu(\bm{x}) = \frac{\exp \left( -\beta E(\bm{x}) \right)}{Z}, \\
Z = \sum_{\bm{x}} \exp \left( -\beta E(\bm{x}) \right),
\end{align}
where $\beta = 1/k_B T$ is the inverse temperature. For simplicity, we set $k_B = 1$ hereafter. The partition function $Z$ is the sum of the Boltzmann factors $\exp(-\beta E(\bm{x}))$ over all spin configurations. However, as the number of spins grows, the state space expands exponentially, rendering the computation of the partition function in large systems extremely difficult.

\subsection{Markov chain Monte Carlo}
\label{sec:Markov_Chain_Monte_Carlo}
MCMC \cite{metropolis1953equation, hastings1970monte} is a widely used method for sampling from complex distributions, including the Boltzmann distribution. 
Starting from a random initial configuration, the algorithm updates this configuration according to a Markov chain with transition probability $P(\bm{x}'|\bm{x})$, ultimately producing samples that follow the target distribution $\pi(\bm{x})$.
The rationale for using Markov chains is that an irreducible and aperiodic chain has a unique stationary distribution, which can be realized by satisfying the detailed balance:
\begin{align}
    \pi(\bm{x}) P(\bm{x}' | \bm{x}) = \pi(\bm{x}') P(\bm{x} | \bm{x}') \quad \forall \bm{x}, \bm{x}'.
\end{align}
The Metropolis-Hastings (MH) algorithm \cite{hastings1970monte} gives a general framework for constructing a Markov chain that meets the detailed balance. 
The central idea is to split the transition probability into a proposal distribution $Q(\bm{x}'|\bm{x})$ and an acceptance probability $A(\bm{x}'|\bm{x})$. 
Specifically, the transition probability is defined as
\begin{align}
P(\bm{x}' | \bm{x}) =
\begin{cases}
Q(\bm{x}'|\bm{x})A(\bm{x}'|\bm{x}) & \text{if } \bm{x}' \neq \bm{x}, \\
1 - \sum_{\bm{x}''\neq\bm{x}} Q(\bm{x}''|\bm{x})A(\bm{x}''|\bm{x}) & \text{if } \bm{x}' = \bm{x},
\end{cases}
\end{align}
where the acceptance probability for a given proposal distribution is defined as
\begin{align}
\label{eq:MH_acceptance}
A(\bm{x}'|\bm{x}) = \min \left( 1, \frac{\pi(\bm{x}')}{\pi(\bm{x})} \frac{Q(\bm{x}|\bm{x}')}{Q(\bm{x}'|\bm{x})} \right).
\end{align}
By accepting proposed transitions with $A(\bm{x}'|\bm{x})$, the chain satisfies detailed balance and converges to the target distribution $\pi(\bm{x})$ after a sufficient number of steps. Notably, the acceptance probability depends only on the ratio of the Boltzmann distributions, thus circumventing the need to evaluate the partition function.

In the MH algorithm, the convergence time depends on the choice of the proposal distribution, making its selection a critical factor.
The single spin-flip (SSF) update, which flips one randomly selected spin, serves as a basic proposal distribution.
Although it is extremely simple and applicable to a wide range of models, it often fails when dealing with distributions characterized by complex energy landscapes.
For example, in systems with strong frustration, such as spin glasses, the energy function includes numerous high-energy barriers.
As a result, the convergence time for low-temperature Boltzmann distributions diverges, effectively breaking the ergodicity of the chain \cite{RevModPhys.58.801}.
To address these challenges, more advanced methods have been proposed for handling complex models.
A prominent example is the cluster update algorithm \cite{swendsen1987nonuniversal, wolff1989collective, houdayer2001cluster}, which simultaneously updates multiple variables according to specific rules, thereby reducing convergence times and enabling more efficient simulations.
However, identifying suitable update rules tailored to a specific model remains highly challenging, and to date, no universal cluster update applicable to all models has been discovered.
Other notable approaches relevant to spin glasses include parallel tempering \cite{PhysRevLett.57.2607, doi:10.1143/JPSJ.65.1604} and the isoenergetic cluster algorithm \cite{zhu2015efficient}.
Despite these advancements, sampling from the Boltzmann distribution in spin glasses remains a major challenge, with many open questions still unresolved.

\subsection{Generative neural sampler}
\label{sec:neural_network}
Neural networks can infer an underlying unknown probability distribution from a given dataset, a task known as distribution estimation that has been studied for decades.
GNSs employ this capability by sampling from the probability distribution learned by these neural networks.
One of the earliest examples in this field is the Boltzmann machine \cite{ACKLEY1985147}. 
In particular, the development of restricted Boltzmann machines (RBMs) \cite{smolensky1986information} and gradient-based methods such as contrastive divergence \cite{hinton2002training} has made it feasible to avoid enumerating exponentially large state spaces, thereby reducing computational costs. 
However, distributions whose normalization constants are hard to compute, such as the Boltzmann distribution, remain challenging for RBMs.
More recently, Neural Autoregressive Distribution Estimators (NADE) \cite{pmlr-v15-larochelle11a, uria2016neural} have proven effective for this task. 
These models offer greater ease of use and higher computational efficiency compared to conventional approaches. 
\begin{figure}
    \centering
    \includegraphics[width=\linewidth]{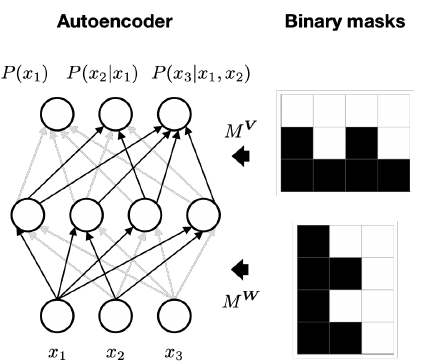}
    \caption{An example of a MADE network with a single hidden layer. By applying a binary mask (shown on the right) that restricts certain connections, the model obtains autoregressive properties absent in a standard autoencoder. The gray paths in MADE represent connections that are removed by the mask.}
    \label{fig:made_scheme}
\end{figure}

In this paper, we focus on the Masked Autoencoder Distribution Estimator (MADE) \cite{germain2015made}, which is a variant of NADE.
MADE is a distribution estimator that uses an autoencoder.
An autoencoder \cite{doi:10.1126/science.1127647} is a feedforward network designed to produce an output vector $\bm{y}$ that closely matches the input vector $\bm{x}$. 
The main objective is to capture the underlying structure of the distribution from which the input data are drawn, thereby learning representations that can efficiently generate new samples from the same distribution. 
An autoencoder generally consists of an input layer, one or more hidden layers, and an output layer. 
A network with a single hidden layer can be written as
\begin{gather}
    \bm{h} = \bm{g} (\bm{W}\bm{x} + \bm{b}) \\
    \bm{y} = {\rm{sigm}}(\bm{V}\bm{h} + \bm{c}),
\end{gather}
where $\bm{W}$ and $\bm{V}$ are weight matrices, $\bm{b}$ and $\bm{c}$ are bias terms, and $\bm{g}$ denotes the activation function. The sigmoid function ${\rm{sigm}}(a) = 1/(1+\exp(-a))$ is applied at the output layer.

MADE augments this autoencoder network with an autoregressive property to facilitate efficient distribution estimation. Any joint distribution can be factorized into a product of conditional probabilities:
\begin{align}
    \label{eq:product_rule}
    p(\bm{x}) = \prod_{d=1}^D p(x_{o_d} | \bm{x}_{o_{<d}})
\end{align}
where $o_d$ specifies the ordering of the variables as a permutation of $\{1, \cdots, D\}$. 
Autoregression imposes dependencies among the output-layer variables, allowing the joint distribution to be computed as the product of the network outputs. 
To implement this, masks $M^{\bm{W}}$ and $M^{\bm{V}}$ are introduced to constrain the network connections. 
Consequently, a single-hidden-layer MADE can be expressed as
\begin{gather}
    \label{eq:made_network_hidden}
    \bm{h} = \bm{g} \left( (M^{\bm{W}} \cdot \bm{W})\bm{x} + \bm{b} \right) \\
    \label{eq:made_network_output}
    \bm{y} = p(x_{o_d}=1 | \bm{x}_{o_{<d}}) = {\rm{sigm}} \left( (M^{\bm{V}} \cdot \bm{V})\bm{h} + \bm{c} \right)
\end{gather}
Figure~\ref{fig:made_scheme} illustrates a case of a MADE network with a single hidden layer and the masks defined by Eqs.~\eqref{eq:made_network_hidden}, \eqref{eq:made_network_output}.
The network is trained by minimizing the cross-entropy loss:
\begin{align}
    l(\bm{x}) = \sum_{d=1}^D \left\{ -x_d \log y_d - (1-x_d) \log (1-y_d) \right\}
\end{align}  
which is equivalent to the negative log-likelihood of $p(\bm{x})$. 
In practice, the loss function can be optimized via stochastic gradient descent, such as Adam \cite{kingma2017adammethodstochasticoptimization}.

Recently, a strategy that integrates machine learning into MCMC has been proposed in the field of statistical physics, involving global updates through GNS that approximate the target distribution \cite{PhysRevE.101.023304,PhysRevE.101.053312}.
These neural proposals enable model-specific global updates by training a GNS to learn an auxiliary distribution that closely approximates the target.
However, a major challenge in this approach lies in constructing a dataset that accurately reflects the target distribution.
In practice, datasets are often generated using MCMC samples obtained via the SSF update.
Yet, neural proposals based on such samples often fail to capture complex energy landscapes with multiple modes, such as those found in low-temperature Boltzmann distributions \cite{Ciarella_2023}.

\subsection{Quantum Approximate Optimization Algorithm}
\label{sec:QAOA}
\begin{figure}
    \centering
    \includegraphics[width=\linewidth]{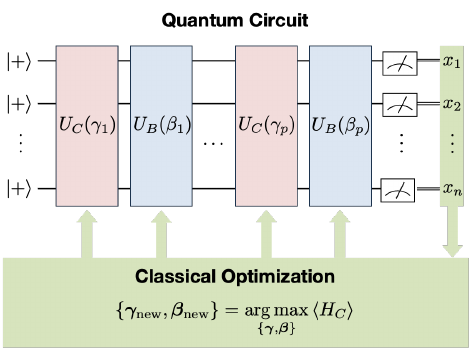}
    \caption{The quantum circuit and algorithm flow of QAOA.
    A parameterized quantum state is generated from the quantum circuit shown at the top.
    Using this state, the energy expectation value is calculated, and the parameters are continuously updated until the energy expectation value is minimized.}
    \label{fig:QAOA}
\end{figure}
QAOA \cite{farhi2014quantum} is a VQA for solving combinatorial optimization problems. 
Its main objective is to find the ground state of a target Hamiltonian $H_C$. 
Specifically, it proceeds as follows: on a quantum computer, a parameterized trial state is generated by a parametrized quantum circuit, and the expectation value of $H_C$ is computed.
On a classical computer, this expectation value is used as the objective function in a nonlinear optimization solver. 
By iterating between these two steps, the algorithm searches for parameters that produce the ground state of $H_C$.

Figure~\ref{fig:QAOA} visualizes the implementation of a $p$-layer QAOA.
The QAOA circuit employs multiple layers, each consisting of alternating operations associated with two different Hamiltonians:
\begin{gather}
    U(\bm{\gamma}, \bm{\beta}) = U_B(\beta_p)U_C(\gamma_p) \cdots U_B(\beta_1)U_C(\gamma_1), 
    \label{eq:qaoa_circuit} \\
    U_B(\beta) = \exp(-i \beta H_B) , \quad U_C(\gamma) = \exp(-i \gamma H_C).
\end{gather}
Here, $H_B$, called the mixer Hamiltonian, does not commute with $H_C$ and is given as:
\begin{align}
    H_B = \sum_{j=1}^n X_j,
\end{align}
where $X_j$ denotes the Pauli-$X$ matrix acting on site $j$. 
Depending on the number of layers $p$, this quantum circuit includes $2p$ adjustable parameters $\{\gamma_j, \beta_j\}_{j=1}^p$.

QAOA was originally designed to solve combinatorial problems, but it can also be used as an “approximate” Boltzmann sampler.  
QAOA is inspired by quantum annealing (QA) \cite{farhi2000quantumcomputationadiabaticevolution, kadowaki1998quantum}, which aims to find the ground state of $H_C$ by continuously evolving the ground state of an initial Hamiltonian over time in accordance with the adiabatic theorem.  
Ideally, QA would guarantee convergence to the ground state.  
In practice, however, finite annealing times and thermal noise cause QA to yield an approximate Boltzmann distribution at a finite temperature \cite{PhysRevApplied.8.064025, PhysRevApplied.17.044046, PRXQuantum.3.020317}.  
Given that QAOA is a discrete approximation of the short-time evolution in QA, it is natural to expect it to produce similar outputs.
Indeed, some numerical evidence indicates that QAOA output distributions similarly approximate a low-temperature Boltzmann distribution \cite{lotshaw2023approximate, diez2023quantum, PhysRevResearch.6.013071}.

\section{QAOA-Trained Neural MCMC}
\label{sec:nmcmc_qaoa}
We present a hybrid quantum-classical MCMC algorithm that integrates GNS with quantum circuits.
In this approach, GNS, which approximates the output distribution of the quantum circuit, serves as the proposal distribution for MCMC.
Specifically, we propose using QAOA for the Boltzmann sampling of spin glasses.

\subsection{Algorithm overview}
The algorithm based on QAOA operates as follows.
First, QAOA is executed, and samples are drawn from its final state.
Although QAOA parameters typically require variational optimization, this computational cost can be significantly reduced or even avoided entirely by using the initial parameters introduced later in this paper.
The resulting binary samples are then used as training data for GNS, allowing it to learn the QAOA output distribution.
Finally, an MCMC simulation is performed using GNS, which approximates the QAOA output distribution, as the proposal distribution.
The full procedure of the proposed method is summarized in Algorithm~\ref{alg:proposed_algorithm}, and a schematic illustration of the overall process is shown in Fig.~\ref{fig:main_scheme}.

The transition defined by GNS is independent of the current state, meaning that $Q(\bm{x}'|\bm{x}) = p(\bm{x}')$.
If $p(\bm{x})$ exactly matches $\pi(\bm{x})$, the acceptance probability becomes 1.
Therefore, when GNS closely approximates the target distribution, a high acceptance rate can be expected.

In this algorithm, MADE is adopted as the GNS.
MADE utilizes autoregressive properties to reduce connections, resulting in polynomial complexity of $O(n^2)$ for both evaluating and sampling from the distribution. 
This structure enables efficient computation.
Furthermore, since the transition is independent of the current state, GNS proposals can be generated separately from the MCMC simulation.
This property suggests that proposals may be precomputed, potentially reducing the overall computational cost.
\begin{algorithm}[H]
    \caption{QAOA-Trained Neural MCMC}
    \label{alg:proposed_algorithm}
    \begin{algorithmic}
    \State $\textbf{Input: } \text{QAOA circuit} \ U(\bm{\theta}),\ \text{parameters}\ \bm{\theta},$
    \State $\text{the dataset size}\ N,\ \text{GNS,}\ \text{the number of MCMC steps}\ M$
    \State $\textbf{Output: } \text{Samples}\ \{\bm{x}^{(m)}\}_{m=1}^M$
    \Statex
        \State $\textbf{Step 1: Sampling from QAOA circuit}$
        \State $\text{Set}\ \bm{\theta}\ \text{in}\ U(\bm{\theta})$
        \State $\text{Dataset}\ \mathcal{D} \leftarrow \text{Sampling} \  U(\bm{\theta}) \  \text{in the computational basis}$
        \Statex
        \State $\textbf{Step 2: Training GNS on QAOA samples}$
        \State $\text{Train GNS on\ } \mathcal{D}$
        \Statex
        \State $\textbf{Step 3: MCMC with trained GNS proposal}$
        \State $\text{Initialize}\ \bm{x}^{(0)}$
        \State $m \leftarrow 0$
        \For {$m < M$}
            \State $\bm{x}',  p(\bm{x}') \leftarrow \text{GNS}$
            \State $A \leftarrow \min \left\{1, \frac{\pi(\bm{x}')}{\pi(\bm{x}^{(m)})}\frac{p(\bm{x}^{(m)})}{p(\bm{x}')} \right\}$
            \If {$A \geq \text{Uniform(0, 1)}$}
                \State ${\bm{x}}^{(m+1)} \leftarrow \bm{x'}$
            \Else
                \State ${\bm{x}}^{(m+1)} \leftarrow \bm{x}^{(m)}$
            \EndIf
            \State $m \leftarrow m+1$
        \EndFor
    \end{algorithmic}
\end{algorithm}

\subsection{Boltzmann Sampling with QAOA-Trained GNS Proposal}
There are several advantages to using QAOA samples as training data for the neural proposal.
The QAOA output distribution is known to approximate a classical Boltzmann distribution at a low temperature, which can be especially beneficial at low temperatures compared to samples generated by traditional MCMC methods such as the SSF update.
In practice, however, the optimized QAOA output distribution may deviate from the true Boltzmann distribution of the model.
This deviation arises because the optimal QAOA solution corresponds to a zero-temperature thermal state, which may fail to capture peaks associated with certain low-energy excited states that are present in a finite-temperature Boltzmann distribution.
Nevertheless, the method often performs well in practical applications.
Indeed, studies have shown that under short-time evolution with QAOA-like Hamiltonians, wavefunction collapse upon measurement tends to converge with high probability to one of the lowest-energy configurations \cite{mazzola2021sampling, layden2023quantum, PhysRevResearch.6.013071}.
This result suggests that the output distribution of a shallow QAOA circuit is well-suited for proposing low-energy configurations, which remains a central challenge in Boltzmann sampling.

Another important feature of QAOA is the availability of heuristic initial parameters. It has been demonstrated that ``fixed angles" \cite{brandao2018fixedcontrolparametersquantum, 9605328, wurtz2021fixedangleconjectureqaoa, 10.1145/3584706} can offer strong performance. These angles depend only on the graph structure and circuit depth, rather than on the system size. They can be used as initial values for parameter optimization, or even directly, as they often yield satisfactory performance on combinatorial optimization problems. Utilizing such parameters enables rapid sample generation via QAOA without the computational overhead of optimization. This feature aligns well with the capabilities of NISQ devices, where expectation value estimation can be challenging, but direct sampling remains feasible. 
As a result, this approach is especially promising for near-term quantum devices.

While this discussion focuses on QAOA implemented on quantum computers, related algorithms using quantum annealers are also possible \cite{10.21468/SciPostPhys.15.1.018}, since QAOA is originally regarded as a discretized form of quantum annealing. 
Quantum annealers are capable of solving Ising models involving thousands of qubits \cite{king2023quantum}, although hardware constraints limit the range of graph structures that can be directly implemented \cite{PRXQuantum.2.040322}. 
In contrast, gate-model quantum computers may support broader applications beyond the Ising model and may not be subject to such graph limitations, depending on the architecture. 
Moreover, VQAs are generally resilient to noise, and several noise mitigation techniques have been developed.
Through these strategies, the proposed method may achieve quantum speedup on real NISQ devices for useful tasks such as Boltzmann sampling.

\section{Numerical experiments}
\label{sec:numerical_experiment}
In this section, we evaluate the performance of the proposed method by sampling from the Boltzmann distribution of spin glass systems.
For all experiments, we assume that computations are carried out on an ideal, noise-free quantum computer.

\subsection{Spectral gap analysis}   
\label{sec:spectral_gap}
We begin by analyzing the average performance of the proposed method for small-scale systems.
Specifically, we calculate the spectral gap of the Markov chains generated by the proposed method and compare these results with those from various alternative proposal distributions.
The spectral gap is defined as the difference between the largest eigenvalue $\lambda_1$ and the second-largest eigenvalue $\lambda_2$ of a Markov chain’s transition probability matrix \cite{levin2017markov}.
Because $P$ is a stochastic matrix, its largest eigenvalue is 1, and the spectral gap is $\delta = 1 - |\lambda_2|$.
This gap is closely linked to the mixing time, which measures the time required for a Markov chain’s distance to the target distribution to be small.
However, for a chain generated via the MH algorithm, the transition matrix is of size $2^n \times 2^n$, where its entries are given by $P(\bm{x}'|\bm{x})$.
Thus, evaluating the eigenvalues of an exponentially large transition matrix becomes infeasible for large-scale systems.

In this experiment, we compute the spectral gap for 100 randomly generated instances for each system size $n$, ranging from 3 to 12 spins, in order to examine how the performance of the proposed method depends on system size.
We consider two types of QAOA circuits: one that uses fixed angles derived from the Sherrington-Kirkpatrick model \cite{basso_et_al:LIPIcs.TQC.2022.7} followed by optimization, and one that applies the same angles without any optimization.
For comparison, we evaluate the proposed method against a quantum algorithm known as QAOA-MC \cite{PhysRevResearch.6.033105}.
This hybrid quantum-classical MCMC algorithm uses a QAOA-inspired circuit as its proposal distribution and adjusts the circuit parameters using a cost function that can be estimated within the MCMC framework to speed up convergence.
We also compare the proposed method with classical proposals, including the SSF update and a uniform proposal distribution.
While more advanced techniques such as replica exchange methods are available for spin glasses, this study limits its scope to a range of simpler proposal distributions.
Importantly, the proposed method could be extended to include replica exchange or other advanced MCMC algorithms for direct comparisons with state-of-the-art methods. 
However, such extensions are beyond the scope of this study.
Detailed settings for QAOA, MADE, and MCMC are provided in Appendix~\ref{sec:numerical_details}.

\begin{figure}
    \centering
    \includegraphics[width=\linewidth]{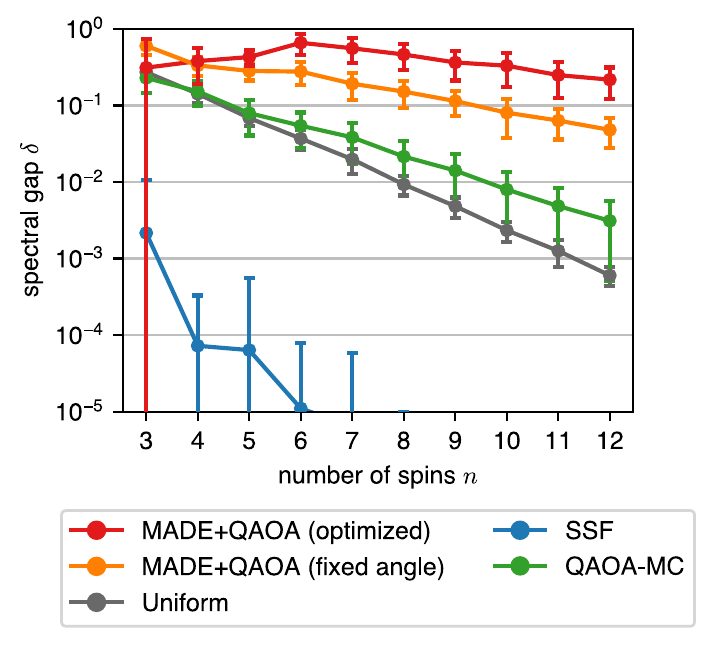}
    \caption{System size dependence of the spectral gap at $\beta = 10$. The data points represent the average spectral gap calculated over 100 randomly generated instances for each system size $n$.}
    \label{fig:spectral_gap}
\end{figure}
Figure~\ref{fig:spectral_gap} shows how the spectral gap varies with $n$ for each proposal distribution at $\beta=10$. 
In this low-temperature regime, the proposed method consistently outperforms all of the other methods. 
Classical proposal distributions struggle to navigate the complex energy landscape, especially near local minima.
By contrast, the proposed approach attains high convergence efficiency thanks to a neural network sampler trained on samples generated by QAOA, which effectively captures low-energy configurations of the spin glass. 
For $n \ge 10$ it delivers a spectral-gap enhancement of roughly two orders of magnitude over a uniform proposal.
Although QAOA-MC also employs QAOA-inspired circuits to facilitate fast transitions between low-energy states, the proposed method still achieves better performance overall.
Unlike QAOA-MC, which is constrained by available quantum circuits, this method, using the GNS representation, provides transitions that are better suited for low-temperature Boltzmann sampling through the QAOA output distribution.

Notably, the fixed-angle QAOA also yields superior outcomes compared to other methods aside from the optimized QAOA.
This finding is particularly significant because the main bottleneck in VQAs generally lies in optimizing the parameterized quantum circuits. 
The possibility of achieving quantum advantage without extensive optimization could thus offer substantial benefits.
Indeed, the hybrid quantum-classical MCMC algorithms typically face optimization and tuning challenges. 
For instance, quantum-enhanced MCMC methods still require careful circuit parameter adjustments \cite{christmann2024quantum}. 
Such difficulties underscore why the proposed method may represent a more practical quantum solution for current NISQ devices.

\subsection{Magnetization estimation}
\label{sec:magnetization}
Next, we assess the performance of the proposed method for larger systems by estimating magnetization. 
Defined as the average spin value in the model, $m(\bm{x}) = \frac{1}{n} \sum_{j=1}^n x_j$, magnetization quantifies the net polarization of spins. 
In this experiment, we estimate the average magnetization $\overline{m} = \sum_{\bm{x}} \mu(\bm{x}) m(\bm{x})$ from samples generated by MCMC simulations of a 25-site spin glass at $\beta=5$, corresponding to the Boltzmann distribution.
We run ten separate MCMC simulations, each starting from a different initial configuration and running for ${10}^5$ steps. 
The QAOA and MADE settings can be found in Appendix~\ref{sec:numerical_details}. 

\begin{figure*}[t]
    \centering
    \includegraphics[width=\linewidth]{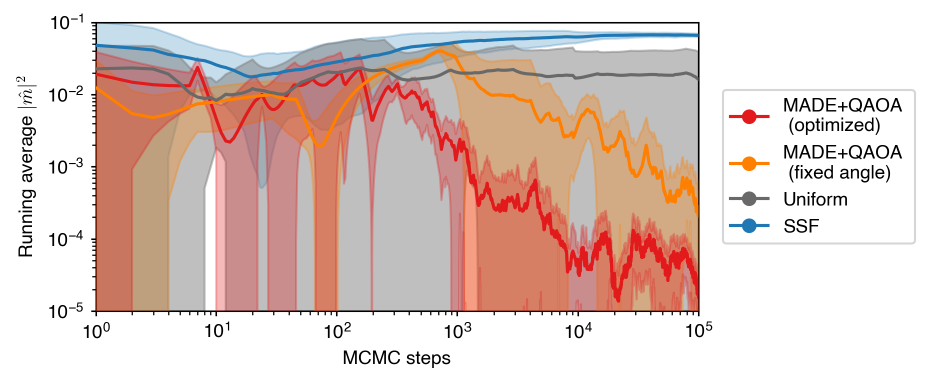}
    \caption{MCMC estimation results for the squared estimator of the average magnetization ($\overline{m}=0$). The solid lines represent the average of running averages, computed from 10 independent Markov chains at each step. The shaded bands around the lines indicate the standard deviations, illustrating variability among the chains.}
    \label{fig:average_magnetization}
\end{figure*}
In Figure~\ref{fig:average_magnetization}, we plot the evolution of the squared estimator of the average magnetization, \(\hat{m}^2 = \left(\frac{1}{M} \sum_{j=0}^{M-1} m({\bm{x}}^{(j)})\right)^2\), as computed from the MCMC simulations.
Because the SSF update only allows local moves, it captures only a limited portion of the energy landscape, causing the magnetization estimate to converge to a different value. 
By contrast, the proposed method steadily approaches the true magnetization. 
Although a uniform proposal permits global transitions, it does so indiscriminately across all states. 
In a low-temperature Boltzmann distribution with sharp energy minima, this uniform search appears to hinder convergence.
In comparison, the proposed method rapidly converges and efficiently estimates the magnetization by prioritizing proposals in low-energy regions of the energy landscape. 
Consequently, it focuses on the most significant configurations for accurate sampling and estimation, showcasing a clear advantage in exploring low-energy states.

\begin{figure*}
    \centering
    \includegraphics[width=\linewidth]{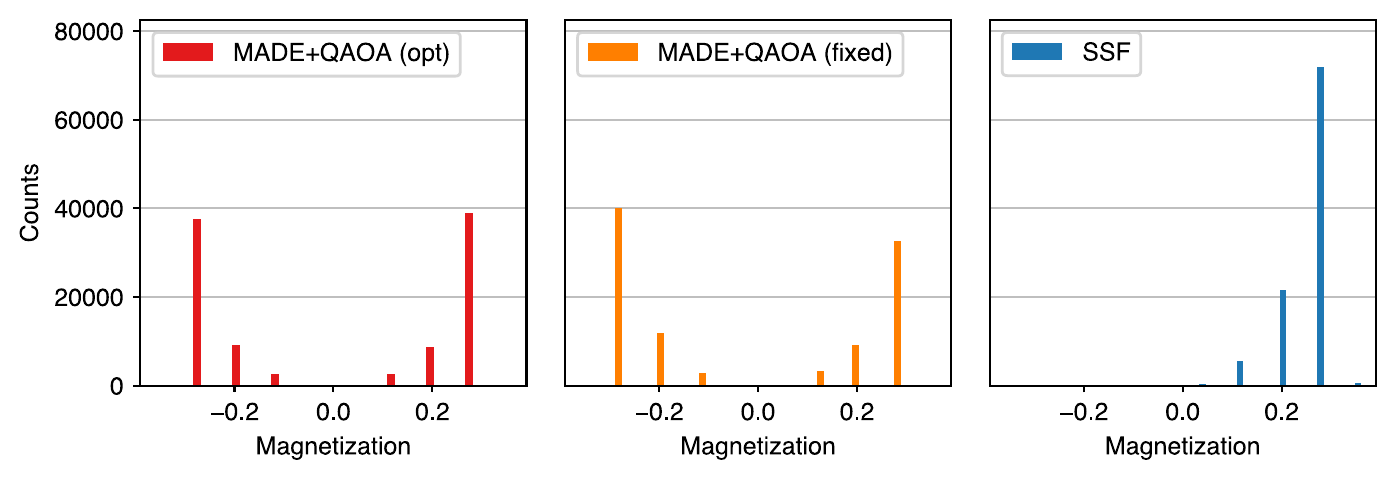}
    \caption{Magnetization histogram for samples obtained from a single MCMC chain.}
    \label{fig:magnetization_histogram}
\end{figure*}
Figure~\ref{fig:magnetization_histogram} presents the magnetization histogram for samples obtained from a single MCMC chain, all starting from the same initial configuration. 
Owing to the $\mathbb{Z}_2$-symmetry of the spin glass defined in Eq.~\eqref{eq:spin_glass}, the system features symmetric magnetization peaks corresponding to pairs of configurations with identical energies.
While the SSF update tends to sample unevenly by favoring one side of these peaks, the proposed method successfully samples both peaks in a balanced manner. 
This highlights its capacity to respect the inherent symmetry and to explore the overall configuration space more thoroughly.

Finally, we examine the autocorrelation of the magnetization. 
The autocorrelation function measures the degree of statistical correlation among MCMC samples. 
Although samples drawn from an ideal probability distribution would be uncorrelated, MCMC samples inherently exhibit finite correlations due to the Markov chain nature of the process. 
When these correlations diminish sufficiently, we regard the MCMC process as converged, making autocorrelation a key metric for gauging convergence efficiency.
The autocorrelation function for magnetization is defined as
\begin{align}
c(\tau) = \frac{\braket{m(t+\tau)m(t)} - {\braket{m(t)}}^2}{\braket{m(t)m(t)} - {\braket{m(t)}}^2},
\end{align}
where $\tau$ is the time shift (lag) in MCMC steps. 
Here, $m(t)$ denotes the magnetization of sample $\bm{x}^{(t)}$, and $\braket{\cdot}$ indicates an average over the entire Markov chain. 
To mitigate biases introduced by initial conditions, we discard the first ${10}^4$ samples of each chain before computing autocorrelations.

\begin{figure}
    \centering
    \includegraphics[width=\linewidth]{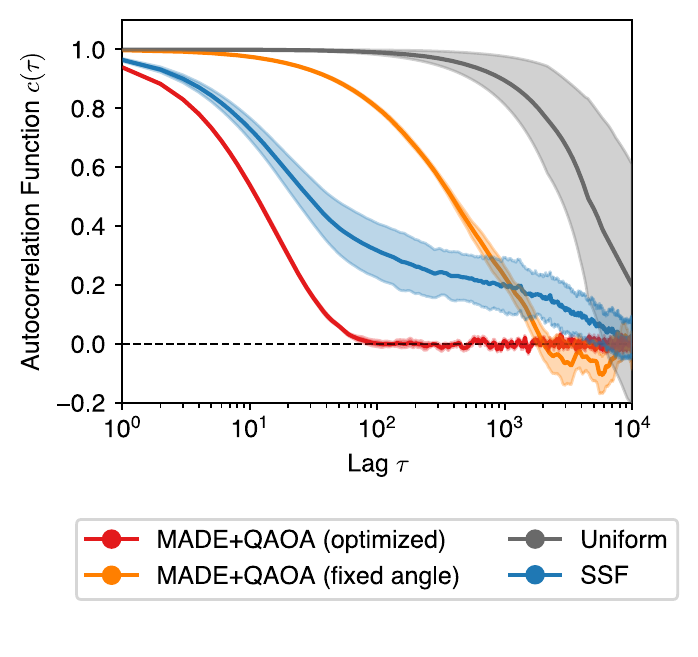}
    \caption{Evolution of the autocorrelation function of magnetization.}
    \label{fig:autocorrelation}
\end{figure}
Figure~\ref{fig:autocorrelation} shows how the magnetization’s autocorrelation evolves over time. 
The proposal distribution derived from optimized QAOA exhibits a notably faster decay in autocorrelation than other methods. 
By contrast, the fixed-angle QAOA, unlike the trend observed in the spectral gap results, performs on par with or worse than the SSF update. 
Nevertheless, it is important to emphasize that the SSF update fails to converge to the correct magnetization, as evidenced by the mean magnetization estimates and magnetization histograms, while the proposed method reliably converges to the true value.
These observations suggest that the proposed method yields an efficient proposal distribution for complex systems such as spin glasses. 
Unlike the SSF update, it achieves unbiased sampling while maintaining ergodicity and successfully capturing the Boltzmann distribution.

\section{Conclusion and outlook}
\label{sec:conclusion}
In this study, we present a method to accelerate MCMC sampling for spin glasses by combining QAOA with a generative network sampler.
While Boltzmann distributions of spin glasses have traditionally been studied in statistical physics, they are also central to binary combinatorial optimization and machine learning.
However, sampling from these distributions remains a well-known challenge, particularly at low temperatures.
The proposed approach leverages samples from a QAOA circuit that approximates the Boltzmann distribution and uses them to train MADE, producing a proposal distribution closely aligned with the target.
We benchmarked the proposed method on fully connected spin glasses and conducted several computational analyses.
Spectral gap results show that the proposed method maintains strong performance as the system size increases.
In magnetization estimation via MCMC, we demonstrated that, unlike the SSF update, the proposed method enables unbiased sampling and converges more rapidly than classical proposals.
This advantage is maintained even when using unoptimized QAOA samples. This finding highlights the potential of the proposed approach to exploit current quantum devices without the extensive measurements that are typically required in VQAs.

An important open question is whether this numerical advantage extends to larger models.
While Ref.~\cite{10.21468/SciPostPhys.15.1.018} conducted experiments on systems with hundreds of sites using a quantum annealer, classical QAOA simulators are not well suited for models at this scale.
Therefore, direct testing on large-scale quantum hardware will be essential to validate the method’s effectiveness.
Moreover, noise in quantum devices could potentially reduce the numerical benefits of the proposed method.
Nonetheless, since hybrid quantum-classical MCMC algorithms affect only the convergence efficiency without violating detailed balance, the proposed method remains promising for near-term quantum computing applications.

Another direction for future research is to extend beyond the Ising model to more general systems.
This extension may require exploring alternative neural network architectures in place of MADE.
Recent studies on neural network samplers for MCMC have indicated that MADE-based proposals may become inefficient for models that involve substantial computational complexity \cite{Ciarella_2023}.
As a result, adopting more advanced architectures such as Transformers, which have shown strong performance in modern machine learning \cite{NIPS2017_3f5ee243, nagai2024self}, or customized architectures tailored to specific physical systems \cite{biazzo2023autoregressive}, could offer significant advantages.

\begin{acknowledgments}
The authors thank Yasuyuki Kato for helpful discussions on the MCMC simulation of spin glasses.
Y.N. is supported by Grant-in-Aid for JSPS Fellows Grant No. 24KJ1606.
K.F. and K.N.O are supported by MEXT Quantum Leap Flagship Program (MEXT Q-LEAP) Grant No. JPMXS0120319794, and JST COI-NEXT Grant No. JPMJPF2014.
K.F. is also supported by JST COI-NEXT Grant No. JPMJPF2014.
\end{acknowledgments}

\appendix

\section{Numerical Details}
\label{sec:numerical_details}
In this section, we describe the detailed settings used in our numerical experiments.
All source code and data are publicly available on the Zenodo repository \cite{nakano_2025_15546116}.

QAOA circuits are sampled at depth $p = 5$, initialized with fixed angles for the Sherrington-Kirkpatrick model \cite{basso_et_al:LIPIcs.TQC.2022.7}.
Quantum circuit simulations are conducted using Qulacs \cite{suzuki2021qulacs}, and circuit parameters are optimized with the BFGS method \cite{byrd1995limited}, as implemented in SciPy \cite{virtanen2020scipy}.

For MADE, we adopt a two-layer hidden architecture.
The construction and execution of the neural network are implemented entirely in PyTorch \cite{NEURIPS2019_bdbca288}.
Table~\ref{tab:parameter_settings} summarizes the MADE settings used in each experiment.
The acceptance rate of the GNS proposal distribution may be affected by the size of the hidden layers and the amount of training data.
If the dataset or latent capacity is insufficient, the network may fail to accurately approximate the output distribution of the quantum circuit, leading to reduced MCMC efficiency \cite{PhysRevE.101.023304, Ciarella_2023}.
In this experiment, these parameters were selected to be sufficiently large, based on several preliminary tests.
It should be noted, however, that all parameters, including these, were selected heuristically, and there remains room for further tuning.
\begin{table*}[htbp]
    \centering
    \caption{Hyperparameters of the MADE model and its training.}
    \begin{tabular}{ccccccccc}
        \toprule
        & Hidden layers & Hidden nodes & Optimizer & Learning rate & Training data & Test data & Batch size & Epochs \\ \hline
        Spectral gap & 2 & $2n$ & Adam & 0.005 & 1000 & 250 & 8 & 30 \\
        Magnetization & 2 & $2n$ & Adam & 0.005 & 8000 & 2000 & 8 & 30 \\
        \bottomrule
    \end{tabular}
    \label{tab:parameter_settings}
\end{table*}

\section{Supplemental Data}
\label{sec:supp_data}
\begin{figure*}
    \centering
    \includegraphics[width=\linewidth]{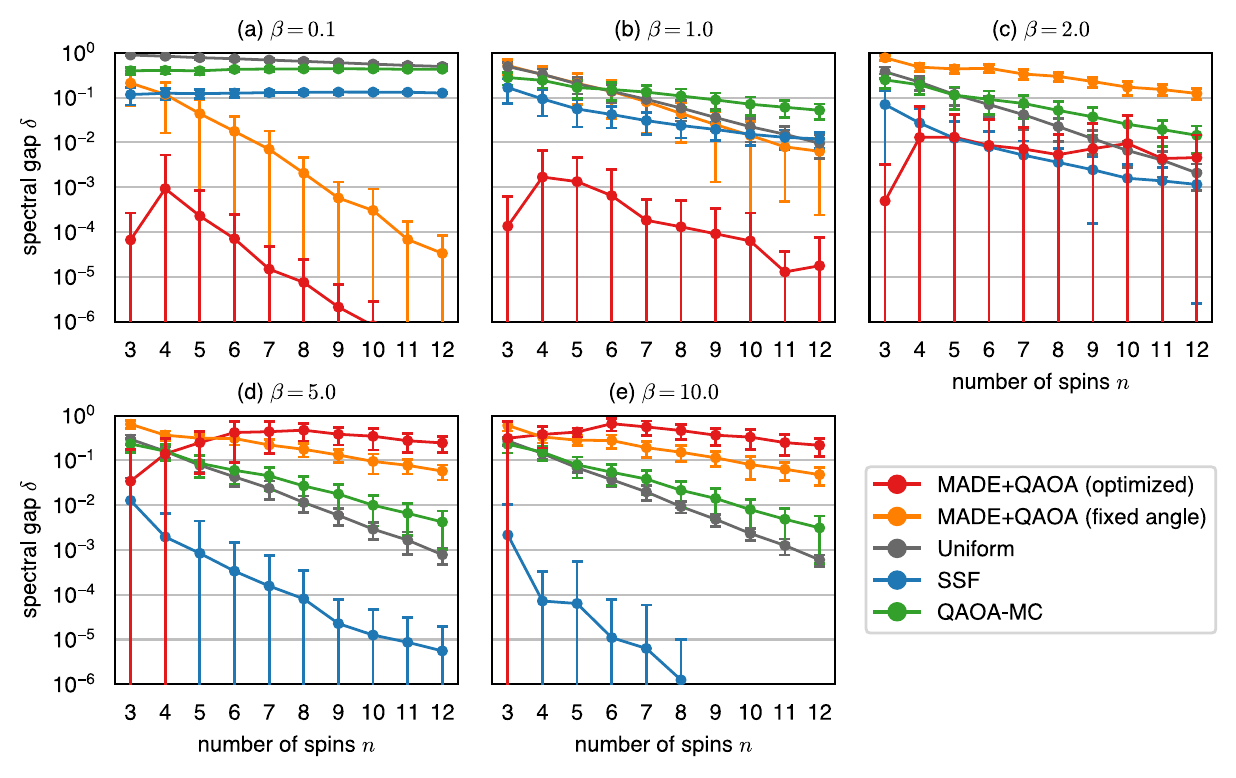}
    \caption{System size dependence of the spectral gap at different temperatures.}
    \label{fig:spectral_gap_appendix}
\end{figure*}
Figure~\ref{fig:spectral_gap_appendix} shows how the spectral gap depends on system size across different temperatures, including the data at $\beta = 10$ previously shown in Fig.~\ref{fig:spectral_gap}.
At high temperatures ($\beta \leq 1$), classical proposals are sufficiently efficient and outperform the proposed method.
It is important to note that simulations in this regime are classically tractable, and the proposed method is specifically designed to utilize approximate Boltzmann distributions generated by QAOA at low temperatures. 
Therefore, it is not intended for high-temperature settings.
In the low-temperature regime ($\beta > 1$), the optimized QAOA circuit yields the best performance.
Interestingly, around $\beta = 1$, the fixed-angle QAOA performs better than its optimized counterpart.
The proposed method consistently maintains an advantage over classical proposals throughout the low-temperature regime.
This result suggests that even without parameter optimization, quantum circuits can serve as effective generators of Boltzmann samples at finite temperatures, which is a highly promising finding.

\bibliographystyle{apsrev4-2}
\bibliography{cite}

\begin{thebibliography}{73}%
\makeatletter
\providecommand \@ifxundefined [1]{%
 \@ifx{#1\undefined}
}%
\providecommand \@ifnum [1]{%
 \ifnum #1\expandafter \@firstoftwo
 \else \expandafter \@secondoftwo
 \fi
}%
\providecommand \@ifx [1]{%
 \ifx #1\expandafter \@firstoftwo
 \else \expandafter \@secondoftwo
 \fi
}%
\providecommand \natexlab [1]{#1}%
\providecommand \enquote  [1]{``#1''}%
\providecommand \bibnamefont  [1]{#1}%
\providecommand \bibfnamefont [1]{#1}%
\providecommand \citenamefont [1]{#1}%
\providecommand \href@noop [0]{\@secondoftwo}%
\providecommand \href [0]{\begingroup \@sanitize@url \@href}%
\providecommand \@href[1]{\@@startlink{#1}\@@href}%
\providecommand \@@href[1]{\endgroup#1\@@endlink}%
\providecommand \@sanitize@url [0]{\catcode `\\12\catcode `\$12\catcode `\&12\catcode `\#12\catcode `\^12\catcode `\_12\catcode `\%12\relax}%
\providecommand \@@startlink[1]{}%
\providecommand \@@endlink[0]{}%
\providecommand \url  [0]{\begingroup\@sanitize@url \@url }%
\providecommand \@url [1]{\endgroup\@href {#1}{\urlprefix }}%
\providecommand \urlprefix  [0]{URL }%
\providecommand \Eprint [0]{\href }%
\providecommand \doibase [0]{https://doi.org/}%
\providecommand \selectlanguage [0]{\@gobble}%
\providecommand \bibinfo  [0]{\@secondoftwo}%
\providecommand \bibfield  [0]{\@secondoftwo}%
\providecommand \translation [1]{[#1]}%
\providecommand \BibitemOpen [0]{}%
\providecommand \bibitemStop [0]{}%
\providecommand \bibitemNoStop [0]{.\EOS\space}%
\providecommand \EOS [0]{\spacefactor3000\relax}%
\providecommand \BibitemShut  [1]{\csname bibitem#1\endcsname}%
\let\auto@bib@innerbib\@empty
\bibitem [{\citenamefont {Shor}(1994)}]{shor1994algorithms}%
  \BibitemOpen
  \bibfield  {author} {\bibinfo {author} {\bibfnamefont {P.~W.}\ \bibnamefont {Shor}},\ }in\ \href {https://doi.org/10.1109/SFCS.1994.365700} {\emph {\bibinfo {booktitle} {Proceedings 35th annual symposium on foundations of computer science}}}\ (\bibinfo {organization} {IEEE},\ \bibinfo {year} {1994})\ pp.\ \bibinfo {pages} {124--134}\BibitemShut {NoStop}%
\bibitem [{\citenamefont {Grover}(1997)}]{grover1997quantum}%
  \BibitemOpen
  \bibfield  {author} {\bibinfo {author} {\bibfnamefont {L.~K.}\ \bibnamefont {Grover}},\ }\href {https://doi.org/10.1103/PhysRevLett.79.325} {\bibfield  {journal} {\bibinfo  {journal} {Phys. Rev. Lett.}\ }\textbf {\bibinfo {volume} {79}},\ \bibinfo {pages} {325} (\bibinfo {year} {1997})}\BibitemShut {NoStop}%
\bibitem [{\citenamefont {Abrams}\ and\ \citenamefont {Lloyd}(1999)}]{abrams1999quantum}%
  \BibitemOpen
  \bibfield  {author} {\bibinfo {author} {\bibfnamefont {D.~S.}\ \bibnamefont {Abrams}}\ and\ \bibinfo {author} {\bibfnamefont {S.}~\bibnamefont {Lloyd}},\ }\href {https://doi.org/10.1103/PhysRevLett.83.5162} {\bibfield  {journal} {\bibinfo  {journal} {Phys. Rev. Lett.}\ }\textbf {\bibinfo {volume} {83}},\ \bibinfo {pages} {5162} (\bibinfo {year} {1999})}\BibitemShut {NoStop}%
\bibitem [{\citenamefont {Harrow}\ \emph {et~al.}(2009)\citenamefont {Harrow}, \citenamefont {Hassidim},\ and\ \citenamefont {Lloyd}}]{PhysRevLett.103.150502}%
  \BibitemOpen
  \bibfield  {author} {\bibinfo {author} {\bibfnamefont {A.~W.}\ \bibnamefont {Harrow}}, \bibinfo {author} {\bibfnamefont {A.}~\bibnamefont {Hassidim}},\ and\ \bibinfo {author} {\bibfnamefont {S.}~\bibnamefont {Lloyd}},\ }\href {https://doi.org/10.1103/PhysRevLett.103.150502} {\bibfield  {journal} {\bibinfo  {journal} {Phys. Rev. Lett.}\ }\textbf {\bibinfo {volume} {103}},\ \bibinfo {pages} {150502} (\bibinfo {year} {2009})}\BibitemShut {NoStop}%
\bibitem [{\citenamefont {Preskill}(2018)}]{preskill2018quantum}%
  \BibitemOpen
  \bibfield  {author} {\bibinfo {author} {\bibfnamefont {J.}~\bibnamefont {Preskill}},\ }\href {https://doi.org/10.22331/q-2018-08-06-79} {\bibfield  {journal} {\bibinfo  {journal} {Quantum}\ }\textbf {\bibinfo {volume} {2}},\ \bibinfo {pages} {79} (\bibinfo {year} {2018})}\BibitemShut {NoStop}%
\bibitem [{\citenamefont {Arute}\ \emph {et~al.}(2019)\citenamefont {Arute}, \citenamefont {Arya}, \citenamefont {Babbush}, \citenamefont {Bacon}, \citenamefont {Bardin}, \citenamefont {Barends}, \citenamefont {Biswas}, \citenamefont {Boixo}, \citenamefont {Brandao}, \citenamefont {Buell} \emph {et~al.}}]{arute2019quantum}%
  \BibitemOpen
  \bibfield  {author} {\bibinfo {author} {\bibfnamefont {F.}~\bibnamefont {Arute}}, \bibinfo {author} {\bibfnamefont {K.}~\bibnamefont {Arya}}, \bibinfo {author} {\bibfnamefont {R.}~\bibnamefont {Babbush}}, \bibinfo {author} {\bibfnamefont {D.}~\bibnamefont {Bacon}}, \bibinfo {author} {\bibfnamefont {J.~C.}\ \bibnamefont {Bardin}}, \bibinfo {author} {\bibfnamefont {R.}~\bibnamefont {Barends}}, \bibinfo {author} {\bibfnamefont {R.}~\bibnamefont {Biswas}}, \bibinfo {author} {\bibfnamefont {S.}~\bibnamefont {Boixo}}, \bibinfo {author} {\bibfnamefont {F.~G.}\ \bibnamefont {Brandao}}, \bibinfo {author} {\bibfnamefont {D.~A.}\ \bibnamefont {Buell}}, \emph {et~al.},\ }\href {https://doi.org/10.1038/s41586-019-1666-5} {\bibfield  {journal} {\bibinfo  {journal} {Nature}\ }\textbf {\bibinfo {volume} {574}},\ \bibinfo {pages} {505} (\bibinfo {year} {2019})}\BibitemShut {NoStop}%
\bibitem [{\citenamefont {Han-Sen}\ \emph {et~al.}(2020)\citenamefont {Han-Sen}, \citenamefont {Hui}, \citenamefont {Yu-Hao}, \citenamefont {Ming-Cheng}, \citenamefont {Li-Chao}, \citenamefont {Yi-Han}, \citenamefont {Jian}, \citenamefont {Dian}, \citenamefont {Xing}, \citenamefont {Yi},\ and\ \citenamefont {others.}}]{doi:10.1126/science.abe8770}%
  \BibitemOpen
  \bibfield  {author} {\bibinfo {author} {\bibfnamefont {Z.}~\bibnamefont {Han-Sen}}, \bibinfo {author} {\bibfnamefont {W.}~\bibnamefont {Hui}}, \bibinfo {author} {\bibfnamefont {D.}~\bibnamefont {Yu-Hao}}, \bibinfo {author} {\bibfnamefont {C.}~\bibnamefont {Ming-Cheng}}, \bibinfo {author} {\bibfnamefont {P.}~\bibnamefont {Li-Chao}}, \bibinfo {author} {\bibfnamefont {L.}~\bibnamefont {Yi-Han}}, \bibinfo {author} {\bibfnamefont {Q.}~\bibnamefont {Jian}}, \bibinfo {author} {\bibfnamefont {W.}~\bibnamefont {Dian}}, \bibinfo {author} {\bibfnamefont {D.}~\bibnamefont {Xing}}, \bibinfo {author} {\bibfnamefont {H.}~\bibnamefont {Yi}},\ and\ \bibinfo {author} {\bibnamefont {others.}},\ }\href {https://doi.org/10.1126/science.abe8770} {\bibfield  {journal} {\bibinfo  {journal} {Science}\ }\textbf {\bibinfo {volume} {370}},\ \bibinfo {pages} {1460} (\bibinfo {year} {2020})},\ \Eprint {https://arxiv.org/abs/https://www.science.org/doi/pdf/10.1126/science.abe8770} {https://www.science.org/doi/pdf/10.1126/science.abe8770}
  \BibitemShut {NoStop}%
\bibitem [{\citenamefont {Madsen}\ \emph {et~al.}(2022)\citenamefont {Madsen}, \citenamefont {Laudenbach}, \citenamefont {Askarani}, \citenamefont {Rortais}, \citenamefont {Vincent}, \citenamefont {Bulmer}, \citenamefont {Miatto}, \citenamefont {Neuhaus}, \citenamefont {Helt}, \citenamefont {Collins},\ and\ \citenamefont {others.}}]{Madsen2022}%
  \BibitemOpen
  \bibfield  {author} {\bibinfo {author} {\bibfnamefont {L.~S.}\ \bibnamefont {Madsen}}, \bibinfo {author} {\bibfnamefont {F.}~\bibnamefont {Laudenbach}}, \bibinfo {author} {\bibfnamefont {M.~F.}\ \bibnamefont {Askarani}}, \bibinfo {author} {\bibfnamefont {F.}~\bibnamefont {Rortais}}, \bibinfo {author} {\bibfnamefont {T.}~\bibnamefont {Vincent}}, \bibinfo {author} {\bibfnamefont {J.~F.~F.}\ \bibnamefont {Bulmer}}, \bibinfo {author} {\bibfnamefont {F.~M.}\ \bibnamefont {Miatto}}, \bibinfo {author} {\bibfnamefont {L.}~\bibnamefont {Neuhaus}}, \bibinfo {author} {\bibfnamefont {L.~G.}\ \bibnamefont {Helt}}, \bibinfo {author} {\bibfnamefont {M.~J.}\ \bibnamefont {Collins}},\ and\ \bibinfo {author} {\bibnamefont {others.}},\ }\href {https://doi.org/10.1038/s41586-022-04725-x} {\bibfield  {journal} {\bibinfo  {journal} {Nature}\ }\textbf {\bibinfo {volume} {606}},\ \bibinfo {pages} {75} (\bibinfo {year} {2022})}\BibitemShut {NoStop}%
\bibitem [{\citenamefont {Cerezo}\ \emph {et~al.}(2021)\citenamefont {Cerezo}, \citenamefont {Arrasmith}, \citenamefont {Babbush}, \citenamefont {Benjamin}, \citenamefont {Endo}, \citenamefont {Fujii}, \citenamefont {McClean}, \citenamefont {Mitarai}, \citenamefont {Yuan}, \citenamefont {Cincio} \emph {et~al.}}]{cerezo2021variational}%
  \BibitemOpen
  \bibfield  {author} {\bibinfo {author} {\bibfnamefont {M.}~\bibnamefont {Cerezo}}, \bibinfo {author} {\bibfnamefont {A.}~\bibnamefont {Arrasmith}}, \bibinfo {author} {\bibfnamefont {R.}~\bibnamefont {Babbush}}, \bibinfo {author} {\bibfnamefont {S.~C.}\ \bibnamefont {Benjamin}}, \bibinfo {author} {\bibfnamefont {S.}~\bibnamefont {Endo}}, \bibinfo {author} {\bibfnamefont {K.}~\bibnamefont {Fujii}}, \bibinfo {author} {\bibfnamefont {J.~R.}\ \bibnamefont {McClean}}, \bibinfo {author} {\bibfnamefont {K.}~\bibnamefont {Mitarai}}, \bibinfo {author} {\bibfnamefont {X.}~\bibnamefont {Yuan}}, \bibinfo {author} {\bibfnamefont {L.}~\bibnamefont {Cincio}}, \emph {et~al.},\ }\href {https://doi.org/10.1038/s42254-021-00348-9} {\bibfield  {journal} {\bibinfo  {journal} {Nat. Rev. Phys.}\ }\textbf {\bibinfo {volume} {3}},\ \bibinfo {pages} {625} (\bibinfo {year} {2021})}\BibitemShut {NoStop}%
\bibitem [{\citenamefont {Farhi}\ \emph {et~al.}(2014)\citenamefont {Farhi}, \citenamefont {Goldstone},\ and\ \citenamefont {Gutmann}}]{farhi2014quantum}%
  \BibitemOpen
  \bibfield  {author} {\bibinfo {author} {\bibfnamefont {E.}~\bibnamefont {Farhi}}, \bibinfo {author} {\bibfnamefont {J.}~\bibnamefont {Goldstone}},\ and\ \bibinfo {author} {\bibfnamefont {S.}~\bibnamefont {Gutmann}},\ }\Eprint {https://arxiv.org/abs/1411.4028} {arXiv:1411.4028 [quant-ph]}  (\bibinfo {year} {2014})\BibitemShut {NoStop}%
\bibitem [{\citenamefont {Peruzzo}\ \emph {et~al.}(2014)\citenamefont {Peruzzo}, \citenamefont {McClean}, \citenamefont {Shadbolt}, \citenamefont {Yung}, \citenamefont {Zhou}, \citenamefont {Love}, \citenamefont {Aspuru-Guzik},\ and\ \citenamefont {O’brien}}]{peruzzo2014variational}%
  \BibitemOpen
  \bibfield  {author} {\bibinfo {author} {\bibfnamefont {A.}~\bibnamefont {Peruzzo}}, \bibinfo {author} {\bibfnamefont {J.}~\bibnamefont {McClean}}, \bibinfo {author} {\bibfnamefont {P.}~\bibnamefont {Shadbolt}}, \bibinfo {author} {\bibfnamefont {M.-H.}\ \bibnamefont {Yung}}, \bibinfo {author} {\bibfnamefont {X.-Q.}\ \bibnamefont {Zhou}}, \bibinfo {author} {\bibfnamefont {P.~J.}\ \bibnamefont {Love}}, \bibinfo {author} {\bibfnamefont {A.}~\bibnamefont {Aspuru-Guzik}},\ and\ \bibinfo {author} {\bibfnamefont {J.~L.}\ \bibnamefont {O’brien}},\ }\href {https://doi.org/10.1038/ncomms5213} {\bibfield  {journal} {\bibinfo  {journal} {Nat. Commun.}\ }\textbf {\bibinfo {volume} {5}},\ \bibinfo {pages} {4213} (\bibinfo {year} {2014})}\BibitemShut {NoStop}%
\bibitem [{\citenamefont {Farhi}\ and\ \citenamefont {Neven}(2018)}]{farhi2018classification}%
  \BibitemOpen
  \bibfield  {author} {\bibinfo {author} {\bibfnamefont {E.}~\bibnamefont {Farhi}}\ and\ \bibinfo {author} {\bibfnamefont {H.}~\bibnamefont {Neven}},\ }\Eprint {https://arxiv.org/abs/1802.06002} {arXiv:1802.06002 [quant-ph]}  (\bibinfo {year} {2018})\BibitemShut {NoStop}%
\bibitem [{\citenamefont {Mitarai}\ \emph {et~al.}(2018)\citenamefont {Mitarai}, \citenamefont {Negoro}, \citenamefont {Kitagawa},\ and\ \citenamefont {Fujii}}]{mitarai2018quantum}%
  \BibitemOpen
  \bibfield  {author} {\bibinfo {author} {\bibfnamefont {K.}~\bibnamefont {Mitarai}}, \bibinfo {author} {\bibfnamefont {M.}~\bibnamefont {Negoro}}, \bibinfo {author} {\bibfnamefont {M.}~\bibnamefont {Kitagawa}},\ and\ \bibinfo {author} {\bibfnamefont {K.}~\bibnamefont {Fujii}},\ }\href {https://doi.org/10.1103/PhysRevA.98.032309} {\bibfield  {journal} {\bibinfo  {journal} {Phys. Rev. A}\ }\textbf {\bibinfo {volume} {98}},\ \bibinfo {pages} {032309} (\bibinfo {year} {2018})}\BibitemShut {NoStop}%
\bibitem [{\citenamefont {Gonthier}\ \emph {et~al.}(2022)\citenamefont {Gonthier}, \citenamefont {Radin}, \citenamefont {Buda}, \citenamefont {Doskocil}, \citenamefont {Abuan},\ and\ \citenamefont {Romero}}]{gonthier2022measurements}%
  \BibitemOpen
  \bibfield  {author} {\bibinfo {author} {\bibfnamefont {J.~F.}\ \bibnamefont {Gonthier}}, \bibinfo {author} {\bibfnamefont {M.~D.}\ \bibnamefont {Radin}}, \bibinfo {author} {\bibfnamefont {C.}~\bibnamefont {Buda}}, \bibinfo {author} {\bibfnamefont {E.~J.}\ \bibnamefont {Doskocil}}, \bibinfo {author} {\bibfnamefont {C.~M.}\ \bibnamefont {Abuan}},\ and\ \bibinfo {author} {\bibfnamefont {J.}~\bibnamefont {Romero}},\ }\href {https://doi.org/10.1103/PhysRevResearch.4.033154} {\bibfield  {journal} {\bibinfo  {journal} {Phys. Rev. Res.}\ }\textbf {\bibinfo {volume} {4}},\ \bibinfo {pages} {033154} (\bibinfo {year} {2022})}\BibitemShut {NoStop}%
\bibitem [{\citenamefont {McClean}\ \emph {et~al.}(2018)\citenamefont {McClean}, \citenamefont {Boixo}, \citenamefont {Smelyanskiy}, \citenamefont {Babbush},\ and\ \citenamefont {Neven}}]{mcclean2018barren}%
  \BibitemOpen
  \bibfield  {author} {\bibinfo {author} {\bibfnamefont {J.~R.}\ \bibnamefont {McClean}}, \bibinfo {author} {\bibfnamefont {S.}~\bibnamefont {Boixo}}, \bibinfo {author} {\bibfnamefont {V.~N.}\ \bibnamefont {Smelyanskiy}}, \bibinfo {author} {\bibfnamefont {R.}~\bibnamefont {Babbush}},\ and\ \bibinfo {author} {\bibfnamefont {H.}~\bibnamefont {Neven}},\ }\href {https://doi.org/10.1038/s41467-018-07090-4} {\bibfield  {journal} {\bibinfo  {journal} {Nature communications}\ }\textbf {\bibinfo {volume} {9}},\ \bibinfo {pages} {4812} (\bibinfo {year} {2018})}\BibitemShut {NoStop}%
\bibitem [{\citenamefont {Hangleiter}\ and\ \citenamefont {Eisert}(2023)}]{hangleiter2023computational}%
  \BibitemOpen
  \bibfield  {author} {\bibinfo {author} {\bibfnamefont {D.}~\bibnamefont {Hangleiter}}\ and\ \bibinfo {author} {\bibfnamefont {J.}~\bibnamefont {Eisert}},\ }\href {https://doi.org/10.1103/RevModPhys.95.035001} {\bibfield  {journal} {\bibinfo  {journal} {Rev. Mod. Phys.}\ }\textbf {\bibinfo {volume} {95}},\ \bibinfo {pages} {035001} (\bibinfo {year} {2023})}\BibitemShut {NoStop}%
\bibitem [{\citenamefont {Aaronson}\ and\ \citenamefont {Chen}(2016)}]{aaronson2016complexity}%
  \BibitemOpen
  \bibfield  {author} {\bibinfo {author} {\bibfnamefont {S.}~\bibnamefont {Aaronson}}\ and\ \bibinfo {author} {\bibfnamefont {L.}~\bibnamefont {Chen}},\ }\href {https://doi.org/10.48550/arXiv.1612.05903} {\bibfield  {journal} {\bibinfo  {journal} {arXiv:1612.05903}\ } (\bibinfo {year} {2016})}\BibitemShut {NoStop}%
\bibitem [{\citenamefont {Kim}\ \emph {et~al.}(2023)\citenamefont {Kim}, \citenamefont {Eddins}, \citenamefont {Anand}, \citenamefont {Wei}, \citenamefont {van~den Berg}, \citenamefont {Rosenblatt}, \citenamefont {Nayfeh}, \citenamefont {Wu}, \citenamefont {Zaletel}, \citenamefont {Temme},\ and\ \citenamefont {Kandala}}]{Kim2023}%
  \BibitemOpen
  \bibfield  {author} {\bibinfo {author} {\bibfnamefont {Y.}~\bibnamefont {Kim}}, \bibinfo {author} {\bibfnamefont {A.}~\bibnamefont {Eddins}}, \bibinfo {author} {\bibfnamefont {S.}~\bibnamefont {Anand}}, \bibinfo {author} {\bibfnamefont {K.~X.}\ \bibnamefont {Wei}}, \bibinfo {author} {\bibfnamefont {E.}~\bibnamefont {van~den Berg}}, \bibinfo {author} {\bibfnamefont {S.}~\bibnamefont {Rosenblatt}}, \bibinfo {author} {\bibfnamefont {H.}~\bibnamefont {Nayfeh}}, \bibinfo {author} {\bibfnamefont {Y.}~\bibnamefont {Wu}}, \bibinfo {author} {\bibfnamefont {M.}~\bibnamefont {Zaletel}}, \bibinfo {author} {\bibfnamefont {K.}~\bibnamefont {Temme}},\ and\ \bibinfo {author} {\bibfnamefont {A.}~\bibnamefont {Kandala}},\ }\href {https://doi.org/10.1038/s41586-023-06096-3} {\bibfield  {journal} {\bibinfo  {journal} {Nature}\ }\textbf {\bibinfo {volume} {618}},\ \bibinfo {pages} {500} (\bibinfo {year} {2023})}\BibitemShut {NoStop}%
\bibitem [{\citenamefont {Kechedzhi}\ \emph {et~al.}(2024)\citenamefont {Kechedzhi}, \citenamefont {Isakov}, \citenamefont {Mandrà}, \citenamefont {Villalonga}, \citenamefont {Mi}, \citenamefont {Boixo},\ and\ \citenamefont {Smelyanskiy}}]{KECHEDZHI2024431}%
  \BibitemOpen
  \bibfield  {author} {\bibinfo {author} {\bibfnamefont {K.}~\bibnamefont {Kechedzhi}}, \bibinfo {author} {\bibfnamefont {S.}~\bibnamefont {Isakov}}, \bibinfo {author} {\bibfnamefont {S.}~\bibnamefont {Mandrà}}, \bibinfo {author} {\bibfnamefont {B.}~\bibnamefont {Villalonga}}, \bibinfo {author} {\bibfnamefont {X.}~\bibnamefont {Mi}}, \bibinfo {author} {\bibfnamefont {S.}~\bibnamefont {Boixo}},\ and\ \bibinfo {author} {\bibfnamefont {V.}~\bibnamefont {Smelyanskiy}},\ }\href {https://doi.org/https://doi.org/10.1016/j.future.2023.12.002} {\bibfield  {journal} {\bibinfo  {journal} {Future Generation Computer Systems}\ }\textbf {\bibinfo {volume} {153}},\ \bibinfo {pages} {431} (\bibinfo {year} {2024})}\BibitemShut {NoStop}%
\bibitem [{\citenamefont {Kanno}\ \emph {et~al.}(2023)\citenamefont {Kanno}, \citenamefont {Kohda}, \citenamefont {Imai}, \citenamefont {Koh}, \citenamefont {Mitarai}, \citenamefont {Mizukami},\ and\ \citenamefont {Nakagawa}}]{kanno2023quantumselectedconfigurationinteractionclassical}%
  \BibitemOpen
  \bibfield  {author} {\bibinfo {author} {\bibfnamefont {K.}~\bibnamefont {Kanno}}, \bibinfo {author} {\bibfnamefont {M.}~\bibnamefont {Kohda}}, \bibinfo {author} {\bibfnamefont {R.}~\bibnamefont {Imai}}, \bibinfo {author} {\bibfnamefont {S.}~\bibnamefont {Koh}}, \bibinfo {author} {\bibfnamefont {K.}~\bibnamefont {Mitarai}}, \bibinfo {author} {\bibfnamefont {W.}~\bibnamefont {Mizukami}},\ and\ \bibinfo {author} {\bibfnamefont {Y.~O.}\ \bibnamefont {Nakagawa}},\ }\Eprint {https://arxiv.org/abs/2302.11320} {arXiv:2302.11320 [quant-ph]}  (\bibinfo {year} {2023})\BibitemShut {NoStop}%
\bibitem [{\citenamefont {Robledo-Moreno}\ \emph {et~al.}(2024)\citenamefont {Robledo-Moreno}, \citenamefont {Motta}, \citenamefont {Haas}, \citenamefont {Javadi-Abhari}, \citenamefont {Jurcevic}, \citenamefont {Kirby}, \citenamefont {Martiel}, \citenamefont {Sharma}, \citenamefont {Sharma}, \citenamefont {Shirakawa},\ and\ \citenamefont {others.}}]{robledomoreno2024chemistryexactsolutionsquantumcentric}%
  \BibitemOpen
  \bibfield  {author} {\bibinfo {author} {\bibfnamefont {J.}~\bibnamefont {Robledo-Moreno}}, \bibinfo {author} {\bibfnamefont {M.}~\bibnamefont {Motta}}, \bibinfo {author} {\bibfnamefont {H.}~\bibnamefont {Haas}}, \bibinfo {author} {\bibfnamefont {A.}~\bibnamefont {Javadi-Abhari}}, \bibinfo {author} {\bibfnamefont {P.}~\bibnamefont {Jurcevic}}, \bibinfo {author} {\bibfnamefont {W.}~\bibnamefont {Kirby}}, \bibinfo {author} {\bibfnamefont {S.}~\bibnamefont {Martiel}}, \bibinfo {author} {\bibfnamefont {K.}~\bibnamefont {Sharma}}, \bibinfo {author} {\bibfnamefont {S.}~\bibnamefont {Sharma}}, \bibinfo {author} {\bibfnamefont {T.}~\bibnamefont {Shirakawa}},\ and\ \bibinfo {author} {\bibnamefont {others.}},\ }\Eprint {https://arxiv.org/abs/2405.05068} {arXiv:2405.05068 [quant-ph]}  (\bibinfo {year} {2024})\BibitemShut {NoStop}%
\bibitem [{\citenamefont {Layden}\ \emph {et~al.}(2023)\citenamefont {Layden}, \citenamefont {Mazzola}, \citenamefont {Mishmash}, \citenamefont {Motta}, \citenamefont {Wocjan}, \citenamefont {Kim},\ and\ \citenamefont {Sheldon}}]{layden2023quantum}%
  \BibitemOpen
  \bibfield  {author} {\bibinfo {author} {\bibfnamefont {D.}~\bibnamefont {Layden}}, \bibinfo {author} {\bibfnamefont {G.}~\bibnamefont {Mazzola}}, \bibinfo {author} {\bibfnamefont {R.~V.}\ \bibnamefont {Mishmash}}, \bibinfo {author} {\bibfnamefont {M.}~\bibnamefont {Motta}}, \bibinfo {author} {\bibfnamefont {P.}~\bibnamefont {Wocjan}}, \bibinfo {author} {\bibfnamefont {J.-S.}\ \bibnamefont {Kim}},\ and\ \bibinfo {author} {\bibfnamefont {S.}~\bibnamefont {Sheldon}},\ }\href {https://doi.org/10.1038/s41586-023-06095-4} {\bibfield  {journal} {\bibinfo  {journal} {Nature}\ }\textbf {\bibinfo {volume} {619}},\ \bibinfo {pages} {282} (\bibinfo {year} {2023})}\BibitemShut {NoStop}%
\bibitem [{\citenamefont {Metropolis}\ \emph {et~al.}(1953)\citenamefont {Metropolis}, \citenamefont {Rosenbluth}, \citenamefont {Rosenbluth}, \citenamefont {Teller},\ and\ \citenamefont {Teller}}]{metropolis1953equation}%
  \BibitemOpen
  \bibfield  {author} {\bibinfo {author} {\bibfnamefont {N.}~\bibnamefont {Metropolis}}, \bibinfo {author} {\bibfnamefont {A.~W.}\ \bibnamefont {Rosenbluth}}, \bibinfo {author} {\bibfnamefont {M.~N.}\ \bibnamefont {Rosenbluth}}, \bibinfo {author} {\bibfnamefont {A.~H.}\ \bibnamefont {Teller}},\ and\ \bibinfo {author} {\bibfnamefont {E.}~\bibnamefont {Teller}},\ }\href {https://doi.org/10.1063/1.1699114} {\bibfield  {journal} {\bibinfo  {journal} {J. Chem. Phys.}\ }\textbf {\bibinfo {volume} {21}},\ \bibinfo {pages} {1087} (\bibinfo {year} {1953})}\BibitemShut {NoStop}%
\bibitem [{\citenamefont {Hastings}(1970)}]{hastings1970monte}%
  \BibitemOpen
  \bibfield  {author} {\bibinfo {author} {\bibfnamefont {W.~K.}\ \bibnamefont {Hastings}},\ }\href {https://doi.org/10.1093/biomet/57.1.97} {\bibfield  {journal} {\bibinfo  {journal} {Biometrika}\ }\textbf {\bibinfo {volume} {57}},\ \bibinfo {pages} {97} (\bibinfo {year} {1970})}\BibitemShut {NoStop}%
\bibitem [{\citenamefont {Nakano}\ \emph {et~al.}(2024)\citenamefont {Nakano}, \citenamefont {Hakoshima}, \citenamefont {Mitarai},\ and\ \citenamefont {Fujii}}]{PhysRevResearch.6.033105}%
  \BibitemOpen
  \bibfield  {author} {\bibinfo {author} {\bibfnamefont {Y.}~\bibnamefont {Nakano}}, \bibinfo {author} {\bibfnamefont {H.}~\bibnamefont {Hakoshima}}, \bibinfo {author} {\bibfnamefont {K.}~\bibnamefont {Mitarai}},\ and\ \bibinfo {author} {\bibfnamefont {K.}~\bibnamefont {Fujii}},\ }\href {https://doi.org/10.1103/PhysRevResearch.6.033105} {\bibfield  {journal} {\bibinfo  {journal} {Phys. Rev. Res.}\ }\textbf {\bibinfo {volume} {6}},\ \bibinfo {pages} {033105} (\bibinfo {year} {2024})}\BibitemShut {NoStop}%
\bibitem [{\citenamefont {Christmann}\ \emph {et~al.}(2024)\citenamefont {Christmann}, \citenamefont {Ivashkov}, \citenamefont {Chiurco},\ and\ \citenamefont {Mazzola}}]{christmann2024quantum}%
  \BibitemOpen
  \bibfield  {author} {\bibinfo {author} {\bibfnamefont {J.}~\bibnamefont {Christmann}}, \bibinfo {author} {\bibfnamefont {P.}~\bibnamefont {Ivashkov}}, \bibinfo {author} {\bibfnamefont {M.}~\bibnamefont {Chiurco}},\ and\ \bibinfo {author} {\bibfnamefont {G.}~\bibnamefont {Mazzola}},\ }\Eprint {https://arxiv.org/abs/2411.17821} {arXiv:2411.17821 [quant-ph]}  (\bibinfo {year} {2024})\BibitemShut {NoStop}%
\bibitem [{\citenamefont {Orfi}\ and\ \citenamefont {Sels}(2024)}]{PhysRevA.110.052414}%
  \BibitemOpen
  \bibfield  {author} {\bibinfo {author} {\bibfnamefont {A.}~\bibnamefont {Orfi}}\ and\ \bibinfo {author} {\bibfnamefont {D.}~\bibnamefont {Sels}},\ }\href {https://doi.org/10.1103/PhysRevA.110.052414} {\bibfield  {journal} {\bibinfo  {journal} {Phys. Rev. A}\ }\textbf {\bibinfo {volume} {110}},\ \bibinfo {pages} {052414} (\bibinfo {year} {2024})}\BibitemShut {NoStop}%
\bibitem [{\citenamefont {Larochelle}\ and\ \citenamefont {Murray}(2011)}]{pmlr-v15-larochelle11a}%
  \BibitemOpen
  \bibfield  {author} {\bibinfo {author} {\bibfnamefont {H.}~\bibnamefont {Larochelle}}\ and\ \bibinfo {author} {\bibfnamefont {I.}~\bibnamefont {Murray}},\ }in\ \href {https://proceedings.mlr.press/v15/larochelle11a.html} {\emph {\bibinfo {booktitle} {Proceedings of the Fourteenth International Conference on Artificial Intelligence and Statistics}}},\ \bibinfo {series} {Proceedings of Machine Learning Research}, Vol.~\bibinfo {volume} {15},\ \bibinfo {editor} {edited by\ \bibinfo {editor} {\bibfnamefont {G.}~\bibnamefont {Gordon}}, \bibinfo {editor} {\bibfnamefont {D.}~\bibnamefont {Dunson}},\ and\ \bibinfo {editor} {\bibfnamefont {M.}~\bibnamefont {Dudík}}}\ (\bibinfo  {publisher} {PMLR},\ \bibinfo {address} {Fort Lauderdale, FL, USA},\ \bibinfo {year} {2011})\ pp.\ \bibinfo {pages} {29--37}\BibitemShut {NoStop}%
\bibitem [{\citenamefont {Uria}\ \emph {et~al.}(2016)\citenamefont {Uria}, \citenamefont {C{{\^o}}t{{\'e}}}, \citenamefont {Gregor}, \citenamefont {Murray},\ and\ \citenamefont {Larochelle}}]{uria2016neural}%
  \BibitemOpen
  \bibfield  {author} {\bibinfo {author} {\bibfnamefont {B.}~\bibnamefont {Uria}}, \bibinfo {author} {\bibfnamefont {M.-A.}\ \bibnamefont {C{{\^o}}t{{\'e}}}}, \bibinfo {author} {\bibfnamefont {K.}~\bibnamefont {Gregor}}, \bibinfo {author} {\bibfnamefont {I.}~\bibnamefont {Murray}},\ and\ \bibinfo {author} {\bibfnamefont {H.}~\bibnamefont {Larochelle}},\ }\href {http://jmlr.org/papers/v17/16-272.html} {\bibfield  {journal} {\bibinfo  {journal} {Journal of Machine Learning Research}\ }\textbf {\bibinfo {volume} {17}},\ \bibinfo {pages} {1} (\bibinfo {year} {2016})}\BibitemShut {NoStop}%
\bibitem [{\citenamefont {Germain}\ \emph {et~al.}(2015)\citenamefont {Germain}, \citenamefont {Gregor}, \citenamefont {Murray},\ and\ \citenamefont {Larochelle}}]{germain2015made}%
  \BibitemOpen
  \bibfield  {author} {\bibinfo {author} {\bibfnamefont {M.}~\bibnamefont {Germain}}, \bibinfo {author} {\bibfnamefont {K.}~\bibnamefont {Gregor}}, \bibinfo {author} {\bibfnamefont {I.}~\bibnamefont {Murray}},\ and\ \bibinfo {author} {\bibfnamefont {H.}~\bibnamefont {Larochelle}},\ }in\ \href {https://proceedings.mlr.press/v37/germain15.html} {\emph {\bibinfo {booktitle} {Proceedings of the 32nd International Conference on Machine Learning}}},\ \bibinfo {series} {Proceedings of Machine Learning Research}, Vol.~\bibinfo {volume} {37},\ \bibinfo {editor} {edited by\ \bibinfo {editor} {\bibfnamefont {F.}~\bibnamefont {Bach}}\ and\ \bibinfo {editor} {\bibfnamefont {D.}~\bibnamefont {Blei}}}\ (\bibinfo  {publisher} {PMLR},\ \bibinfo {address} {Lille, France},\ \bibinfo {year} {2015})\ pp.\ \bibinfo {pages} {881--889}\BibitemShut {NoStop}%
\bibitem [{\citenamefont {Lotshaw}\ \emph {et~al.}(2023)\citenamefont {Lotshaw}, \citenamefont {Siopsis}, \citenamefont {Ostrowski}, \citenamefont {Herrman}, \citenamefont {Alam}, \citenamefont {Powers},\ and\ \citenamefont {Humble}}]{lotshaw2023approximate}%
  \BibitemOpen
  \bibfield  {author} {\bibinfo {author} {\bibfnamefont {P.~C.}\ \bibnamefont {Lotshaw}}, \bibinfo {author} {\bibfnamefont {G.}~\bibnamefont {Siopsis}}, \bibinfo {author} {\bibfnamefont {J.}~\bibnamefont {Ostrowski}}, \bibinfo {author} {\bibfnamefont {R.}~\bibnamefont {Herrman}}, \bibinfo {author} {\bibfnamefont {R.}~\bibnamefont {Alam}}, \bibinfo {author} {\bibfnamefont {S.}~\bibnamefont {Powers}},\ and\ \bibinfo {author} {\bibfnamefont {T.~S.}\ \bibnamefont {Humble}},\ }\href {https://doi.org/10.1103/PhysRevA.108.042411} {\bibfield  {journal} {\bibinfo  {journal} {Phys. Rev. A}\ }\textbf {\bibinfo {volume} {108}},\ \bibinfo {pages} {042411} (\bibinfo {year} {2023})}\BibitemShut {NoStop}%
\bibitem [{\citenamefont {D\'{\i}ez-Valle}\ \emph {et~al.}(2023)\citenamefont {D\'{\i}ez-Valle}, \citenamefont {Porras},\ and\ \citenamefont {Garc\'{\i}a-Ripoll}}]{diez2023quantum}%
  \BibitemOpen
  \bibfield  {author} {\bibinfo {author} {\bibfnamefont {P.}~\bibnamefont {D\'{\i}ez-Valle}}, \bibinfo {author} {\bibfnamefont {D.}~\bibnamefont {Porras}},\ and\ \bibinfo {author} {\bibfnamefont {J.~J.}\ \bibnamefont {Garc\'{\i}a-Ripoll}},\ }\href {https://doi.org/10.1103/PhysRevLett.130.050601} {\bibfield  {journal} {\bibinfo  {journal} {Phys. Rev. Lett.}\ }\textbf {\bibinfo {volume} {130}},\ \bibinfo {pages} {050601} (\bibinfo {year} {2023})}\BibitemShut {NoStop}%
\bibitem [{\citenamefont {Leontica}\ and\ \citenamefont {Amaro}(2024)}]{PhysRevResearch.6.013071}%
  \BibitemOpen
  \bibfield  {author} {\bibinfo {author} {\bibfnamefont {S.}~\bibnamefont {Leontica}}\ and\ \bibinfo {author} {\bibfnamefont {D.}~\bibnamefont {Amaro}},\ }\href {https://doi.org/10.1103/PhysRevResearch.6.013071} {\bibfield  {journal} {\bibinfo  {journal} {Phys. Rev. Res.}\ }\textbf {\bibinfo {volume} {6}},\ \bibinfo {pages} {013071} (\bibinfo {year} {2024})}\BibitemShut {NoStop}%
\bibitem [{\citenamefont {Farhi}\ and\ \citenamefont {Harrow}(2019)}]{farhi2019quantumsupremacyquantumapproximate}%
  \BibitemOpen
  \bibfield  {author} {\bibinfo {author} {\bibfnamefont {E.}~\bibnamefont {Farhi}}\ and\ \bibinfo {author} {\bibfnamefont {A.~W.}\ \bibnamefont {Harrow}},\ }\Eprint {https://arxiv.org/abs/1602.07674} {arXiv:1602.07674 [quant-ph]}  (\bibinfo {year} {2019})\BibitemShut {NoStop}%
\bibitem [{\citenamefont {Brandao}\ \emph {et~al.}(2018)\citenamefont {Brandao}, \citenamefont {Broughton}, \citenamefont {Farhi}, \citenamefont {Gutmann},\ and\ \citenamefont {Neven}}]{brandao2018fixedcontrolparametersquantum}%
  \BibitemOpen
  \bibfield  {author} {\bibinfo {author} {\bibfnamefont {F.~G. S.~L.}\ \bibnamefont {Brandao}}, \bibinfo {author} {\bibfnamefont {M.}~\bibnamefont {Broughton}}, \bibinfo {author} {\bibfnamefont {E.}~\bibnamefont {Farhi}}, \bibinfo {author} {\bibfnamefont {S.}~\bibnamefont {Gutmann}},\ and\ \bibinfo {author} {\bibfnamefont {H.}~\bibnamefont {Neven}},\ }\Eprint {https://arxiv.org/abs/1812.04170} {arXiv:1812.04170 [quant-ph]}  (\bibinfo {year} {2018})\BibitemShut {NoStop}%
\bibitem [{\citenamefont {Galda}\ \emph {et~al.}(2021)\citenamefont {Galda}, \citenamefont {Liu}, \citenamefont {Lykov}, \citenamefont {Alexeev},\ and\ \citenamefont {Safro}}]{9605328}%
  \BibitemOpen
  \bibfield  {author} {\bibinfo {author} {\bibfnamefont {A.}~\bibnamefont {Galda}}, \bibinfo {author} {\bibfnamefont {X.}~\bibnamefont {Liu}}, \bibinfo {author} {\bibfnamefont {D.}~\bibnamefont {Lykov}}, \bibinfo {author} {\bibfnamefont {Y.}~\bibnamefont {Alexeev}},\ and\ \bibinfo {author} {\bibfnamefont {I.}~\bibnamefont {Safro}},\ }in\ \href {https://doi.org/10.1109/QCE52317.2021.00034} {\emph {\bibinfo {booktitle} {2021 IEEE International Conference on Quantum Computing and Engineering (QCE)}}}\ (\bibinfo {year} {2021})\ pp.\ \bibinfo {pages} {171--180}\BibitemShut {NoStop}%
\bibitem [{\citenamefont {Wurtz}\ and\ \citenamefont {Lykov}(2021)}]{wurtz2021fixedangleconjectureqaoa}%
  \BibitemOpen
  \bibfield  {author} {\bibinfo {author} {\bibfnamefont {J.}~\bibnamefont {Wurtz}}\ and\ \bibinfo {author} {\bibfnamefont {D.}~\bibnamefont {Lykov}},\ }\Eprint {https://arxiv.org/abs/2107.00677} {arXiv:2107.00677 [quant-ph]}  (\bibinfo {year} {2021})\BibitemShut {NoStop}%
\bibitem [{\citenamefont {Shaydulin}\ \emph {et~al.}(2023)\citenamefont {Shaydulin}, \citenamefont {Lotshaw}, \citenamefont {Larson}, \citenamefont {Ostrowski},\ and\ \citenamefont {Humble}}]{10.1145/3584706}%
  \BibitemOpen
  \bibfield  {author} {\bibinfo {author} {\bibfnamefont {R.}~\bibnamefont {Shaydulin}}, \bibinfo {author} {\bibfnamefont {P.~C.}\ \bibnamefont {Lotshaw}}, \bibinfo {author} {\bibfnamefont {J.}~\bibnamefont {Larson}}, \bibinfo {author} {\bibfnamefont {J.}~\bibnamefont {Ostrowski}},\ and\ \bibinfo {author} {\bibfnamefont {T.~S.}\ \bibnamefont {Humble}},\ }\href {https://doi.org/10.1145/3584706} {\bibfield  {journal} {\bibinfo  {journal} {ACM Transactions on Quantum Computing}\ }\textbf {\bibinfo {volume} {4}} (\bibinfo {year} {2023})}\BibitemShut {NoStop}%
\bibitem [{\citenamefont {Parisi}(1992)}]{doi:10.1142/1655}%
  \BibitemOpen
  \bibfield  {author} {\bibinfo {author} {\bibfnamefont {G.}~\bibnamefont {Parisi}},\ }\href {https://doi.org/10.1142/1655} {\emph {\bibinfo {title} {Field Theory, Disorder and Simulations}}}\ (\bibinfo  {publisher} {WORLD SCIENTIFIC},\ \bibinfo {year} {1992})\BibitemShut {NoStop}%
\bibitem [{\citenamefont {Binder}\ and\ \citenamefont {Young}(1986)}]{RevModPhys.58.801}%
  \BibitemOpen
  \bibfield  {author} {\bibinfo {author} {\bibfnamefont {K.}~\bibnamefont {Binder}}\ and\ \bibinfo {author} {\bibfnamefont {A.~P.}\ \bibnamefont {Young}},\ }\href {https://doi.org/10.1103/RevModPhys.58.801} {\bibfield  {journal} {\bibinfo  {journal} {Rev. Mod. Phys.}\ }\textbf {\bibinfo {volume} {58}},\ \bibinfo {pages} {801} (\bibinfo {year} {1986})}\BibitemShut {NoStop}%
\bibitem [{\citenamefont {Swendsen}\ and\ \citenamefont {Wang}(1987)}]{swendsen1987nonuniversal}%
  \BibitemOpen
  \bibfield  {author} {\bibinfo {author} {\bibfnamefont {R.~H.}\ \bibnamefont {Swendsen}}\ and\ \bibinfo {author} {\bibfnamefont {J.-S.}\ \bibnamefont {Wang}},\ }\href {https://doi.org/10.1103/PhysRevLett.58.86} {\bibfield  {journal} {\bibinfo  {journal} {Phys. Rev. Lett.}\ }\textbf {\bibinfo {volume} {58}},\ \bibinfo {pages} {86} (\bibinfo {year} {1987})}\BibitemShut {NoStop}%
\bibitem [{\citenamefont {Wolff}(1989)}]{wolff1989collective}%
  \BibitemOpen
  \bibfield  {author} {\bibinfo {author} {\bibfnamefont {U.}~\bibnamefont {Wolff}},\ }\href {https://doi.org/10.1103/PhysRevLett.62.361} {\bibfield  {journal} {\bibinfo  {journal} {Phys. Rev. Lett.}\ }\textbf {\bibinfo {volume} {62}},\ \bibinfo {pages} {361} (\bibinfo {year} {1989})}\BibitemShut {NoStop}%
\bibitem [{\citenamefont {Houdayer}(2001)}]{houdayer2001cluster}%
  \BibitemOpen
  \bibfield  {author} {\bibinfo {author} {\bibfnamefont {J.}~\bibnamefont {Houdayer}},\ }\href {https://doi.org/10.1007/PL00011151} {\bibfield  {journal} {\bibinfo  {journal} {Eur. Phys. J. B}\ }\textbf {\bibinfo {volume} {22}},\ \bibinfo {pages} {479} (\bibinfo {year} {2001})}\BibitemShut {NoStop}%
\bibitem [{\citenamefont {Swendsen}\ and\ \citenamefont {Wang}(1986)}]{PhysRevLett.57.2607}%
  \BibitemOpen
  \bibfield  {author} {\bibinfo {author} {\bibfnamefont {R.~H.}\ \bibnamefont {Swendsen}}\ and\ \bibinfo {author} {\bibfnamefont {J.-S.}\ \bibnamefont {Wang}},\ }\href {https://doi.org/10.1103/PhysRevLett.57.2607} {\bibfield  {journal} {\bibinfo  {journal} {Phys. Rev. Lett.}\ }\textbf {\bibinfo {volume} {57}},\ \bibinfo {pages} {2607} (\bibinfo {year} {1986})}\BibitemShut {NoStop}%
\bibitem [{\citenamefont {Hukushima}\ and\ \citenamefont {Nemoto}(1996)}]{doi:10.1143/JPSJ.65.1604}%
  \BibitemOpen
  \bibfield  {author} {\bibinfo {author} {\bibfnamefont {K.}~\bibnamefont {Hukushima}}\ and\ \bibinfo {author} {\bibfnamefont {K.}~\bibnamefont {Nemoto}},\ }\href {https://doi.org/10.1143/JPSJ.65.1604} {\bibfield  {journal} {\bibinfo  {journal} {Journal of the Physical Society of Japan}\ }\textbf {\bibinfo {volume} {65}},\ \bibinfo {pages} {1604} (\bibinfo {year} {1996})}\BibitemShut {NoStop}%
\bibitem [{\citenamefont {Zhu}\ \emph {et~al.}(2015)\citenamefont {Zhu}, \citenamefont {Ochoa},\ and\ \citenamefont {Katzgraber}}]{zhu2015efficient}%
  \BibitemOpen
  \bibfield  {author} {\bibinfo {author} {\bibfnamefont {Z.}~\bibnamefont {Zhu}}, \bibinfo {author} {\bibfnamefont {A.~J.}\ \bibnamefont {Ochoa}},\ and\ \bibinfo {author} {\bibfnamefont {H.~G.}\ \bibnamefont {Katzgraber}},\ }\href {https://doi.org/10.1103/PhysRevLett.115.077201} {\bibfield  {journal} {\bibinfo  {journal} {Phys. Rev. Lett.}\ }\textbf {\bibinfo {volume} {115}},\ \bibinfo {pages} {077201} (\bibinfo {year} {2015})}\BibitemShut {NoStop}%
\bibitem [{\citenamefont {Ackley}\ \emph {et~al.}(1985)\citenamefont {Ackley}, \citenamefont {Hinton},\ and\ \citenamefont {Sejnowski}}]{ACKLEY1985147}%
  \BibitemOpen
  \bibfield  {author} {\bibinfo {author} {\bibfnamefont {D.~H.}\ \bibnamefont {Ackley}}, \bibinfo {author} {\bibfnamefont {G.~E.}\ \bibnamefont {Hinton}},\ and\ \bibinfo {author} {\bibfnamefont {T.~J.}\ \bibnamefont {Sejnowski}},\ }\href {https://doi.org/https://doi.org/10.1016/S0364-0213(85)80012-4} {\bibfield  {journal} {\bibinfo  {journal} {Cognitive Science}\ }\textbf {\bibinfo {volume} {9}},\ \bibinfo {pages} {147} (\bibinfo {year} {1985})}\BibitemShut {NoStop}%
\bibitem [{\citenamefont {Smolensky}(1986)}]{smolensky1986information}%
  \BibitemOpen
  \bibfield  {author} {\bibinfo {author} {\bibfnamefont {P.}~\bibnamefont {Smolensky}},\ }in\ \href {https://doi.org/10.7551/mitpress/5236.003.0009} {\emph {\bibinfo {booktitle} {Parallel Distributed Processing, Volume 1: Explorations in the Microstructure of Cognition: Foundations}}}\ (\bibinfo  {publisher} {The MIT Press},\ \bibinfo {year} {1986})\BibitemShut {NoStop}%
\bibitem [{\citenamefont {Hinton}(2002)}]{hinton2002training}%
  \BibitemOpen
  \bibfield  {author} {\bibinfo {author} {\bibfnamefont {G.~E.}\ \bibnamefont {Hinton}},\ }\href {https://doi.org/10.1162/089976602760128018} {\bibfield  {journal} {\bibinfo  {journal} {Neural Computation}\ }\textbf {\bibinfo {volume} {14}},\ \bibinfo {pages} {1771} (\bibinfo {year} {2002})}\BibitemShut {NoStop}%
\bibitem [{\citenamefont {Hinton}\ and\ \citenamefont {Salakhutdinov}(2006)}]{doi:10.1126/science.1127647}%
  \BibitemOpen
  \bibfield  {author} {\bibinfo {author} {\bibfnamefont {G.~E.}\ \bibnamefont {Hinton}}\ and\ \bibinfo {author} {\bibfnamefont {R.~R.}\ \bibnamefont {Salakhutdinov}},\ }\href {https://doi.org/10.1126/science.1127647} {\bibfield  {journal} {\bibinfo  {journal} {Science}\ }\textbf {\bibinfo {volume} {313}},\ \bibinfo {pages} {504} (\bibinfo {year} {2006})}\BibitemShut {NoStop}%
\bibitem [{\citenamefont {Kingma}\ and\ \citenamefont {Ba}(2017)}]{kingma2017adammethodstochasticoptimization}%
  \BibitemOpen
  \bibfield  {author} {\bibinfo {author} {\bibfnamefont {D.~P.}\ \bibnamefont {Kingma}}\ and\ \bibinfo {author} {\bibfnamefont {J.}~\bibnamefont {Ba}},\ }\Eprint {https://arxiv.org/abs/1412.6980} {arXiv:1412.6980 [cs.LG]}  (\bibinfo {year} {2017})\BibitemShut {NoStop}%
\bibitem [{\citenamefont {Nicoli}\ \emph {et~al.}(2020)\citenamefont {Nicoli}, \citenamefont {Nakajima}, \citenamefont {Strodthoff}, \citenamefont {Samek}, \citenamefont {M\"uller},\ and\ \citenamefont {Kessel}}]{PhysRevE.101.023304}%
  \BibitemOpen
  \bibfield  {author} {\bibinfo {author} {\bibfnamefont {K.~A.}\ \bibnamefont {Nicoli}}, \bibinfo {author} {\bibfnamefont {S.}~\bibnamefont {Nakajima}}, \bibinfo {author} {\bibfnamefont {N.}~\bibnamefont {Strodthoff}}, \bibinfo {author} {\bibfnamefont {W.}~\bibnamefont {Samek}}, \bibinfo {author} {\bibfnamefont {K.-R.}\ \bibnamefont {M\"uller}},\ and\ \bibinfo {author} {\bibfnamefont {P.}~\bibnamefont {Kessel}},\ }\href {https://doi.org/10.1103/PhysRevE.101.023304} {\bibfield  {journal} {\bibinfo  {journal} {Phys. Rev. E}\ }\textbf {\bibinfo {volume} {101}},\ \bibinfo {pages} {023304} (\bibinfo {year} {2020})}\BibitemShut {NoStop}%
\bibitem [{\citenamefont {McNaughton}\ \emph {et~al.}(2020)\citenamefont {McNaughton}, \citenamefont {Milo\ifmmode \check{s}\else \v{s}\fi{}evi\ifmmode~\acute{c}\else \'{c}\fi{}}, \citenamefont {Perali},\ and\ \citenamefont {Pilati}}]{PhysRevE.101.053312}%
  \BibitemOpen
  \bibfield  {author} {\bibinfo {author} {\bibfnamefont {B.}~\bibnamefont {McNaughton}}, \bibinfo {author} {\bibfnamefont {M.~V.}\ \bibnamefont {Milo\ifmmode \check{s}\else \v{s}\fi{}evi\ifmmode~\acute{c}\else \'{c}\fi{}}}, \bibinfo {author} {\bibfnamefont {A.}~\bibnamefont {Perali}},\ and\ \bibinfo {author} {\bibfnamefont {S.}~\bibnamefont {Pilati}},\ }\href {https://doi.org/10.1103/PhysRevE.101.053312} {\bibfield  {journal} {\bibinfo  {journal} {Phys. Rev. E}\ }\textbf {\bibinfo {volume} {101}},\ \bibinfo {pages} {053312} (\bibinfo {year} {2020})}\BibitemShut {NoStop}%
\bibitem [{\citenamefont {Ciarella}\ \emph {et~al.}(2023)\citenamefont {Ciarella}, \citenamefont {Trinquier}, \citenamefont {Weigt},\ and\ \citenamefont {Zamponi}}]{Ciarella_2023}%
  \BibitemOpen
  \bibfield  {author} {\bibinfo {author} {\bibfnamefont {S.}~\bibnamefont {Ciarella}}, \bibinfo {author} {\bibfnamefont {J.}~\bibnamefont {Trinquier}}, \bibinfo {author} {\bibfnamefont {M.}~\bibnamefont {Weigt}},\ and\ \bibinfo {author} {\bibfnamefont {F.}~\bibnamefont {Zamponi}},\ }\href {https://doi.org/10.1088/2632-2153/acbe91} {\bibfield  {journal} {\bibinfo  {journal} {Machine Learning: Science and Technology}\ }\textbf {\bibinfo {volume} {4}},\ \bibinfo {pages} {010501} (\bibinfo {year} {2023})}\BibitemShut {NoStop}%
\bibitem [{\citenamefont {Farhi}\ \emph {et~al.}(2000)\citenamefont {Farhi}, \citenamefont {Goldstone}, \citenamefont {Gutmann},\ and\ \citenamefont {Sipser}}]{farhi2000quantumcomputationadiabaticevolution}%
  \BibitemOpen
  \bibfield  {author} {\bibinfo {author} {\bibfnamefont {E.}~\bibnamefont {Farhi}}, \bibinfo {author} {\bibfnamefont {J.}~\bibnamefont {Goldstone}}, \bibinfo {author} {\bibfnamefont {S.}~\bibnamefont {Gutmann}},\ and\ \bibinfo {author} {\bibfnamefont {M.}~\bibnamefont {Sipser}},\ }\Eprint {https://arxiv.org/abs/quant-ph/0001106} {arXiv:quant-ph/0001106 [quant-ph]}  (\bibinfo {year} {2000})\BibitemShut {NoStop}%
\bibitem [{\citenamefont {Kadowaki}\ and\ \citenamefont {Nishimori}(1998)}]{kadowaki1998quantum}%
  \BibitemOpen
  \bibfield  {author} {\bibinfo {author} {\bibfnamefont {T.}~\bibnamefont {Kadowaki}}\ and\ \bibinfo {author} {\bibfnamefont {H.}~\bibnamefont {Nishimori}},\ }\href {https://doi.org/10.1103/PhysRevE.58.5355} {\bibfield  {journal} {\bibinfo  {journal} {Physical Review E}\ }\textbf {\bibinfo {volume} {58}},\ \bibinfo {pages} {5355} (\bibinfo {year} {1998})}\BibitemShut {NoStop}%
\bibitem [{\citenamefont {Marshall}\ \emph {et~al.}(2017)\citenamefont {Marshall}, \citenamefont {Rieffel},\ and\ \citenamefont {Hen}}]{PhysRevApplied.8.064025}%
  \BibitemOpen
  \bibfield  {author} {\bibinfo {author} {\bibfnamefont {J.}~\bibnamefont {Marshall}}, \bibinfo {author} {\bibfnamefont {E.~G.}\ \bibnamefont {Rieffel}},\ and\ \bibinfo {author} {\bibfnamefont {I.}~\bibnamefont {Hen}},\ }\href {https://doi.org/10.1103/PhysRevApplied.8.064025} {\bibfield  {journal} {\bibinfo  {journal} {Phys. Rev. Appl.}\ }\textbf {\bibinfo {volume} {8}},\ \bibinfo {pages} {064025} (\bibinfo {year} {2017})}\BibitemShut {NoStop}%
\bibitem [{\citenamefont {Nelson}\ \emph {et~al.}(2022)\citenamefont {Nelson}, \citenamefont {Vuffray}, \citenamefont {Lokhov}, \citenamefont {Albash},\ and\ \citenamefont {Coffrin}}]{PhysRevApplied.17.044046}%
  \BibitemOpen
  \bibfield  {author} {\bibinfo {author} {\bibfnamefont {J.}~\bibnamefont {Nelson}}, \bibinfo {author} {\bibfnamefont {M.}~\bibnamefont {Vuffray}}, \bibinfo {author} {\bibfnamefont {A.~Y.}\ \bibnamefont {Lokhov}}, \bibinfo {author} {\bibfnamefont {T.}~\bibnamefont {Albash}},\ and\ \bibinfo {author} {\bibfnamefont {C.}~\bibnamefont {Coffrin}},\ }\href {https://doi.org/10.1103/PhysRevApplied.17.044046} {\bibfield  {journal} {\bibinfo  {journal} {Phys. Rev. Appl.}\ }\textbf {\bibinfo {volume} {17}},\ \bibinfo {pages} {044046} (\bibinfo {year} {2022})}\BibitemShut {NoStop}%
\bibitem [{\citenamefont {Vuffray}\ \emph {et~al.}(2022)\citenamefont {Vuffray}, \citenamefont {Coffrin}, \citenamefont {Kharkov},\ and\ \citenamefont {Lokhov}}]{PRXQuantum.3.020317}%
  \BibitemOpen
  \bibfield  {author} {\bibinfo {author} {\bibfnamefont {M.}~\bibnamefont {Vuffray}}, \bibinfo {author} {\bibfnamefont {C.}~\bibnamefont {Coffrin}}, \bibinfo {author} {\bibfnamefont {Y.~A.}\ \bibnamefont {Kharkov}},\ and\ \bibinfo {author} {\bibfnamefont {A.~Y.}\ \bibnamefont {Lokhov}},\ }\href {https://doi.org/10.1103/PRXQuantum.3.020317} {\bibfield  {journal} {\bibinfo  {journal} {PRX Quantum}\ }\textbf {\bibinfo {volume} {3}},\ \bibinfo {pages} {020317} (\bibinfo {year} {2022})}\BibitemShut {NoStop}%
\bibitem [{\citenamefont {Mazzola}(2021)}]{mazzola2021sampling}%
  \BibitemOpen
  \bibfield  {author} {\bibinfo {author} {\bibfnamefont {G.}~\bibnamefont {Mazzola}},\ }\href {https://doi.org/10.1103/PhysRevA.104.022431} {\bibfield  {journal} {\bibinfo  {journal} {Physical Review A}\ }\textbf {\bibinfo {volume} {104}},\ \bibinfo {pages} {022431} (\bibinfo {year} {2021})}\BibitemShut {NoStop}%
\bibitem [{\citenamefont {Scriva}\ \emph {et~al.}(2023)\citenamefont {Scriva}, \citenamefont {Costa}, \citenamefont {McNaughton},\ and\ \citenamefont {Pilati}}]{10.21468/SciPostPhys.15.1.018}%
  \BibitemOpen
  \bibfield  {author} {\bibinfo {author} {\bibfnamefont {G.}~\bibnamefont {Scriva}}, \bibinfo {author} {\bibfnamefont {E.}~\bibnamefont {Costa}}, \bibinfo {author} {\bibfnamefont {B.}~\bibnamefont {McNaughton}},\ and\ \bibinfo {author} {\bibfnamefont {S.}~\bibnamefont {Pilati}},\ }\href {https://doi.org/10.21468/SciPostPhys.15.1.018} {\bibfield  {journal} {\bibinfo  {journal} {SciPost Phys.}\ }\textbf {\bibinfo {volume} {15}},\ \bibinfo {pages} {018} (\bibinfo {year} {2023})}\BibitemShut {NoStop}%
\bibitem [{\citenamefont {King}\ \emph {et~al.}(2023)\citenamefont {King}, \citenamefont {Raymond}, \citenamefont {Lanting}, \citenamefont {Harris}, \citenamefont {Zucca}, \citenamefont {Altomare}, \citenamefont {Berkley}, \citenamefont {Boothby}, \citenamefont {Ejtemaee}, \citenamefont {Enderud} \emph {et~al.}}]{king2023quantum}%
  \BibitemOpen
  \bibfield  {author} {\bibinfo {author} {\bibfnamefont {A.~D.}\ \bibnamefont {King}}, \bibinfo {author} {\bibfnamefont {J.}~\bibnamefont {Raymond}}, \bibinfo {author} {\bibfnamefont {T.}~\bibnamefont {Lanting}}, \bibinfo {author} {\bibfnamefont {R.}~\bibnamefont {Harris}}, \bibinfo {author} {\bibfnamefont {A.}~\bibnamefont {Zucca}}, \bibinfo {author} {\bibfnamefont {F.}~\bibnamefont {Altomare}}, \bibinfo {author} {\bibfnamefont {A.~J.}\ \bibnamefont {Berkley}}, \bibinfo {author} {\bibfnamefont {K.}~\bibnamefont {Boothby}}, \bibinfo {author} {\bibfnamefont {S.}~\bibnamefont {Ejtemaee}}, \bibinfo {author} {\bibfnamefont {C.}~\bibnamefont {Enderud}}, \emph {et~al.},\ }\href {https://doi.org/10.1038/s41586-023-05867-2} {\bibfield  {journal} {\bibinfo  {journal} {Nature}\ }\textbf {\bibinfo {volume} {617}},\ \bibinfo {pages} {61} (\bibinfo {year} {2023})}\BibitemShut {NoStop}%
\bibitem [{\citenamefont {K\"onz}\ \emph {et~al.}(2021)\citenamefont {K\"onz}, \citenamefont {Lechner}, \citenamefont {Katzgraber},\ and\ \citenamefont {Troyer}}]{PRXQuantum.2.040322}%
  \BibitemOpen
  \bibfield  {author} {\bibinfo {author} {\bibfnamefont {M.~S.}\ \bibnamefont {K\"onz}}, \bibinfo {author} {\bibfnamefont {W.}~\bibnamefont {Lechner}}, \bibinfo {author} {\bibfnamefont {H.~G.}\ \bibnamefont {Katzgraber}},\ and\ \bibinfo {author} {\bibfnamefont {M.}~\bibnamefont {Troyer}},\ }\href {https://doi.org/10.1103/PRXQuantum.2.040322} {\bibfield  {journal} {\bibinfo  {journal} {PRX Quantum}\ }\textbf {\bibinfo {volume} {2}},\ \bibinfo {pages} {040322} (\bibinfo {year} {2021})}\BibitemShut {NoStop}%
\bibitem [{\citenamefont {Levin}\ and\ \citenamefont {Peres}(2017)}]{levin2017markov}%
  \BibitemOpen
  \bibfield  {author} {\bibinfo {author} {\bibfnamefont {D.~A.}\ \bibnamefont {Levin}}\ and\ \bibinfo {author} {\bibfnamefont {Y.}~\bibnamefont {Peres}},\ }\href@noop {} {\emph {\bibinfo {title} {Markov chains and mixing times}}},\ \bibinfo {edition} {2nd}\ ed.\ (\bibinfo  {publisher} {American Mathematical Soc.},\ \bibinfo {year} {2017})\BibitemShut {NoStop}%
\bibitem [{\citenamefont {Basso}\ \emph {et~al.}(2022)\citenamefont {Basso}, \citenamefont {Farhi}, \citenamefont {Marwaha}, \citenamefont {Villalonga},\ and\ \citenamefont {Zhou}}]{basso_et_al:LIPIcs.TQC.2022.7}%
  \BibitemOpen
  \bibfield  {author} {\bibinfo {author} {\bibfnamefont {J.}~\bibnamefont {Basso}}, \bibinfo {author} {\bibfnamefont {E.}~\bibnamefont {Farhi}}, \bibinfo {author} {\bibfnamefont {K.}~\bibnamefont {Marwaha}}, \bibinfo {author} {\bibfnamefont {B.}~\bibnamefont {Villalonga}},\ and\ \bibinfo {author} {\bibfnamefont {L.}~\bibnamefont {Zhou}},\ }in\ \href {https://drops.dagstuhl.de/entities/document/10.4230/LIPIcs.TQC.2022.7} {\emph {\bibinfo {booktitle} {17th Conference on the Theory of Quantum Computation, Communication and Cryptography (TQC 2022)}}},\ \bibinfo {series} {Leibniz International Proceedings in Informatics (LIPIcs)}, Vol.\ \bibinfo {volume} {232},\ \bibinfo {editor} {edited by\ \bibinfo {editor} {\bibfnamefont {F.}~\bibnamefont {Le~Gall}}\ and\ \bibinfo {editor} {\bibfnamefont {T.}~\bibnamefont {Morimae}}}\ (\bibinfo  {publisher} {Schloss Dagstuhl -- Leibniz-Zentrum f{\"u}r Informatik},\ \bibinfo {address} {Dagstuhl, Germany},\ \bibinfo {year} {2022})\ pp.\ \bibinfo {pages} {7:1--7:21}\BibitemShut
  {NoStop}%
\bibitem [{\citenamefont {Vaswani}\ \emph {et~al.}(2017)\citenamefont {Vaswani}, \citenamefont {Shazeer}, \citenamefont {Parmar}, \citenamefont {Uszkoreit}, \citenamefont {Jones}, \citenamefont {Gomez}, \citenamefont {Kaiser},\ and\ \citenamefont {Polosukhin}}]{NIPS2017_3f5ee243}%
  \BibitemOpen
  \bibfield  {author} {\bibinfo {author} {\bibfnamefont {A.}~\bibnamefont {Vaswani}}, \bibinfo {author} {\bibfnamefont {N.}~\bibnamefont {Shazeer}}, \bibinfo {author} {\bibfnamefont {N.}~\bibnamefont {Parmar}}, \bibinfo {author} {\bibfnamefont {J.}~\bibnamefont {Uszkoreit}}, \bibinfo {author} {\bibfnamefont {L.}~\bibnamefont {Jones}}, \bibinfo {author} {\bibfnamefont {A.~N.}\ \bibnamefont {Gomez}}, \bibinfo {author} {\bibfnamefont {L.~u.}\ \bibnamefont {Kaiser}},\ and\ \bibinfo {author} {\bibfnamefont {I.}~\bibnamefont {Polosukhin}},\ }in\ \href {https://proceedings.neurips.cc/paper_files/paper/2017/file/3f5ee243547dee91fbd053c1c4a845aa-Paper.pdf} {\emph {\bibinfo {booktitle} {Advances in Neural Information Processing Systems}}},\ Vol.~\bibinfo {volume} {30},\ \bibinfo {editor} {edited by\ \bibinfo {editor} {\bibfnamefont {I.}~\bibnamefont {Guyon}}, \bibinfo {editor} {\bibfnamefont {U.~V.}\ \bibnamefont {Luxburg}}, \bibinfo {editor} {\bibfnamefont {S.}~\bibnamefont {Bengio}}, \bibinfo {editor} {\bibfnamefont
  {H.}~\bibnamefont {Wallach}}, \bibinfo {editor} {\bibfnamefont {R.}~\bibnamefont {Fergus}}, \bibinfo {editor} {\bibfnamefont {S.}~\bibnamefont {Vishwanathan}},\ and\ \bibinfo {editor} {\bibfnamefont {R.}~\bibnamefont {Garnett}}}\ (\bibinfo  {publisher} {Curran Associates, Inc.},\ \bibinfo {year} {2017})\BibitemShut {NoStop}%
\bibitem [{\citenamefont {Nagai}\ and\ \citenamefont {Tomiya}(2024)}]{nagai2024self}%
  \BibitemOpen
  \bibfield  {author} {\bibinfo {author} {\bibfnamefont {Y.}~\bibnamefont {Nagai}}\ and\ \bibinfo {author} {\bibfnamefont {A.}~\bibnamefont {Tomiya}},\ }\href {https://doi.org/10.7566/JPSJ.93.114007} {\bibfield  {journal} {\bibinfo  {journal} {Journal of the Physical Society of Japan}\ }\textbf {\bibinfo {volume} {93}},\ \bibinfo {pages} {114007} (\bibinfo {year} {2024})}\BibitemShut {NoStop}%
\bibitem [{\citenamefont {Biazzo}(2023)}]{biazzo2023autoregressive}%
  \BibitemOpen
  \bibfield  {author} {\bibinfo {author} {\bibfnamefont {I.}~\bibnamefont {Biazzo}},\ }\href {https://doi.org/10.1038/s42005-023-01416-5} {\bibfield  {journal} {\bibinfo  {journal} {Communications Physics}\ }\textbf {\bibinfo {volume} {6}},\ \bibinfo {pages} {296} (\bibinfo {year} {2023})}\BibitemShut {NoStop}%
\bibitem [{\citenamefont {Nakano}(2025)}]{nakano_2025_15546116}%
  \BibitemOpen
  \bibfield  {author} {\bibinfo {author} {\bibfnamefont {Y.}~\bibnamefont {Nakano}},\ }\href {https://doi.org/10.5281/zenodo.15546116} {\bibinfo {title} {data\_for\_qaoa-made-mcmc}} (\bibinfo {year} {2025})\BibitemShut {NoStop}%
\bibitem [{\citenamefont {Suzuki}\ \emph {et~al.}(2021)\citenamefont {Suzuki}, \citenamefont {Kawase}, \citenamefont {Masumura}, \citenamefont {Hiraga}, \citenamefont {Nakadai}, \citenamefont {Chen}, \citenamefont {Nakanishi}, \citenamefont {Mitarai}, \citenamefont {Imai}, \citenamefont {Tamiya} \emph {et~al.}}]{suzuki2021qulacs}%
  \BibitemOpen
  \bibfield  {author} {\bibinfo {author} {\bibfnamefont {Y.}~\bibnamefont {Suzuki}}, \bibinfo {author} {\bibfnamefont {Y.}~\bibnamefont {Kawase}}, \bibinfo {author} {\bibfnamefont {Y.}~\bibnamefont {Masumura}}, \bibinfo {author} {\bibfnamefont {Y.}~\bibnamefont {Hiraga}}, \bibinfo {author} {\bibfnamefont {M.}~\bibnamefont {Nakadai}}, \bibinfo {author} {\bibfnamefont {J.}~\bibnamefont {Chen}}, \bibinfo {author} {\bibfnamefont {K.~M.}\ \bibnamefont {Nakanishi}}, \bibinfo {author} {\bibfnamefont {K.}~\bibnamefont {Mitarai}}, \bibinfo {author} {\bibfnamefont {R.}~\bibnamefont {Imai}}, \bibinfo {author} {\bibfnamefont {S.}~\bibnamefont {Tamiya}}, \emph {et~al.},\ }\href {https://doi.org/10.22331/q-2021-10-06-559} {\bibfield  {journal} {\bibinfo  {journal} {Quantum}\ }\textbf {\bibinfo {volume} {5}},\ \bibinfo {pages} {559} (\bibinfo {year} {2021})}\BibitemShut {NoStop}%
\bibitem [{\citenamefont {Byrd}\ \emph {et~al.}(1995)\citenamefont {Byrd}, \citenamefont {Lu}, \citenamefont {Nocedal},\ and\ \citenamefont {Zhu}}]{byrd1995limited}%
  \BibitemOpen
  \bibfield  {author} {\bibinfo {author} {\bibfnamefont {R.~H.}\ \bibnamefont {Byrd}}, \bibinfo {author} {\bibfnamefont {P.}~\bibnamefont {Lu}}, \bibinfo {author} {\bibfnamefont {J.}~\bibnamefont {Nocedal}},\ and\ \bibinfo {author} {\bibfnamefont {C.}~\bibnamefont {Zhu}},\ }\href {https://doi.org/10.1137/0916069} {\bibfield  {journal} {\bibinfo  {journal} {SIAM J. Sci. Comput.}\ }\textbf {\bibinfo {volume} {16}},\ \bibinfo {pages} {1190} (\bibinfo {year} {1995})}\BibitemShut {NoStop}%
\bibitem [{\citenamefont {Virtanen}\ \emph {et~al.}(2020)\citenamefont {Virtanen}, \citenamefont {Gommers}, \citenamefont {Oliphant}, \citenamefont {Haberland}, \citenamefont {Reddy}, \citenamefont {Cournapeau}, \citenamefont {Burovski}, \citenamefont {Peterson}, \citenamefont {Weckesser}, \citenamefont {Bright} \emph {et~al.}}]{virtanen2020scipy}%
  \BibitemOpen
  \bibfield  {author} {\bibinfo {author} {\bibfnamefont {P.}~\bibnamefont {Virtanen}}, \bibinfo {author} {\bibfnamefont {R.}~\bibnamefont {Gommers}}, \bibinfo {author} {\bibfnamefont {T.~E.}\ \bibnamefont {Oliphant}}, \bibinfo {author} {\bibfnamefont {M.}~\bibnamefont {Haberland}}, \bibinfo {author} {\bibfnamefont {T.}~\bibnamefont {Reddy}}, \bibinfo {author} {\bibfnamefont {D.}~\bibnamefont {Cournapeau}}, \bibinfo {author} {\bibfnamefont {E.}~\bibnamefont {Burovski}}, \bibinfo {author} {\bibfnamefont {P.}~\bibnamefont {Peterson}}, \bibinfo {author} {\bibfnamefont {W.}~\bibnamefont {Weckesser}}, \bibinfo {author} {\bibfnamefont {J.}~\bibnamefont {Bright}}, \emph {et~al.},\ }\href {https://doi.org/10.1038/s41592-019-0686-2} {\bibfield  {journal} {\bibinfo  {journal} {Nat. Methods}\ }\textbf {\bibinfo {volume} {17}},\ \bibinfo {pages} {261} (\bibinfo {year} {2020})}\BibitemShut {NoStop}%
\bibitem [{\citenamefont {Paszke}\ \emph {et~al.}(2019)\citenamefont {Paszke}, \citenamefont {Gross}, \citenamefont {Massa}, \citenamefont {Lerer}, \citenamefont {Bradbury}, \citenamefont {Chanan}, \citenamefont {Killeen}, \citenamefont {Lin}, \citenamefont {Gimelshein}, \citenamefont {Antiga} \emph {et~al.}}]{NEURIPS2019_bdbca288}%
  \BibitemOpen
  \bibfield  {author} {\bibinfo {author} {\bibfnamefont {A.}~\bibnamefont {Paszke}}, \bibinfo {author} {\bibfnamefont {S.}~\bibnamefont {Gross}}, \bibinfo {author} {\bibfnamefont {F.}~\bibnamefont {Massa}}, \bibinfo {author} {\bibfnamefont {A.}~\bibnamefont {Lerer}}, \bibinfo {author} {\bibfnamefont {J.}~\bibnamefont {Bradbury}}, \bibinfo {author} {\bibfnamefont {G.}~\bibnamefont {Chanan}}, \bibinfo {author} {\bibfnamefont {T.}~\bibnamefont {Killeen}}, \bibinfo {author} {\bibfnamefont {Z.}~\bibnamefont {Lin}}, \bibinfo {author} {\bibfnamefont {N.}~\bibnamefont {Gimelshein}}, \bibinfo {author} {\bibfnamefont {L.}~\bibnamefont {Antiga}}, \emph {et~al.},\ }in\ \href {https://proceedings.neurips.cc/paper_files/paper/2019/file/bdbca288fee7f92f2bfa9f7012727740-Paper.pdf} {\emph {\bibinfo {booktitle} {Advances in Neural Information Processing Systems}}},\ Vol.~\bibinfo {volume} {32},\ \bibinfo {editor} {edited by\ \bibinfo {editor} {\bibfnamefont {H.}~\bibnamefont {Wallach}}, \bibinfo {editor} {\bibfnamefont
  {H.}~\bibnamefont {Larochelle}}, \bibinfo {editor} {\bibfnamefont {A.}~\bibnamefont {Beygelzimer}}, \bibinfo {editor} {\bibfnamefont {F.}~\bibnamefont {d\textquotesingle Alch\'{e}-Buc}}, \bibinfo {editor} {\bibfnamefont {E.}~\bibnamefont {Fox}},\ and\ \bibinfo {editor} {\bibfnamefont {R.}~\bibnamefont {Garnett}}}\ (\bibinfo  {publisher} {Curran Associates, Inc.},\ \bibinfo {year} {2019})\BibitemShut {NoStop}%
\end{thebibliography}%

\end{document}